\documentclass[useAMS,usenatbib,usegraphicx]{mn2e}
\usepackage{aas_macros,times}

\title[ROLES II: method and SFRD]
  {The Redshift One LDSS-3 Emission line Survey (ROLES) II: \\
  Survey method and z$\sim$1 mass-dependent star-formation rate density}
\author[D. G. Gilbank et al.]
  {David G.~Gilbank$^1$\thanks{Email: dgilbank@astro.uwaterloo.ca}, Michael L.~Balogh$^1$, 
  Karl~Glazebrook$^2$, Richard G.~Bower$^3$,
  \newauthor
  I.K.~Baldry$^4$,  G.T.~Davies$^3$,  
  G.K.T.~Hau$^{2}$, I.H.~Li$^2$, P. ~McCarthy$^5$
\newauthor\\
  $^1$Department of Physics and Astronomy, University of Waterloo, Waterloo, Ontario, Canada N2L 3G1\\
  $^2$Centre for Astrophysics and Supercomputing, Swinburne University of Technology, P.O. Box 218, Hawthorn, VIC 3122, Australia\\
    $^3$Institute for Computational Cosmology, Department of Physics, University of Durham, South Road, Durham, DH1 3LE, UK\\
  $^4$Astrophysics Research Institute, Liverpool John Moores University, Twelve Quays House, Egerton Wharf, Birkenhead CH41 1LD, UK\\
  $^5$Carnegie Observatories, 813 Santa Barbara Street, Pasadena, California, 91101 USA\\
}

\pagerange{\pageref{firstpage}--\pageref{lastpage}} \pubyear{2006}

\def\LaTeX{L\kern-.36em\raise.3ex\hbox{a}\kern-.15em
    T\kern-.1667em\lower.7ex\hbox{E}\kern-.125emX}

\def\oii{[{\sc O\,II}]}
\def\oiii{[{\sc O\,III}]}
\def\nii{[{\sc N\,II}]}
\def\sii{[{\sc S\,II}]}
\def\g09{G10}

\def\ms{$\log(M_*/M_\odot)$}
\def\ha{{\rm H$\alpha$}}

\def\kms{km s$^{-1}$}

\def\lsim{\mathrel{\hbox{\rlap{\hbox{\lower4pt\hbox{$\sim$}}}\hbox{$<$}}}}
\def\gsim{\mathrel{\hbox{\rlap{\hbox{\lower4pt\hbox{$\sim$}}}\hbox{$>$}}}}

\begin{document}

\label{firstpage}

\maketitle

\begin{abstract}
Motivated by suggestions of `cosmic downsizing', in which the dominant contribution to the cosmic star formation rate density (SFRD) proceeds from higher to lower mass galaxies with increasing cosmic time, we describe the design and implementation of the Redshift One LDSS3 Emission line Survey (ROLES). This survey is designed to probe low mass, z$\sim$1 galaxies  directly for the first time with spectroscopy.  ROLES is a $K$-selected ($22.5 < K_{\rm AB} < 24.0$) survey for dwarf galaxies [$8.5 \lsim$\ms$\lsim9.5$] at $0.89 < z < 1.15$ drawn from two extremely deep fields (GOODS-S and MS1054-FIRES).  

Using the \oii$\lambda3727$ emission line, we obtain redshifts and star-formation rates (SFRs) for star-forming galaxies down to a limit of $\sim0.3M_\odot yr^{-1}$.  We present the \oii~luminosity function measured in ROLES and find a faint end slope of $\alpha_{faint}\sim-1.5$, similar to that measured at z$\sim$0.1 in the SDSS. By combining ROLES with higher mass surveys (GDDS and ESO GOOD-S public spectroscopy) we measure the SFRD as a function of stellar mass using \oii~(with and without various empirical corrections), and using SED-fitting to obtain the SFR from the rest-frame UV luminosity for galaxies with spectroscopic redshifts.  Our best estimate of the corrected \oii-SFRD and UV SFRD both independently show that the SFRD evolves equally for galaxies of all masses between z$\sim$1 and z$\sim$0.1.  The exact evolution in normalisation depends on the indicator used, with the \oii-based estimate showing a change of a factor of $\approx$2.6 and the UV-based a factor of $\approx$6.  We discuss possible reasons for the discrepancy in normalisation between the indicators, but note that the magnitude of this uncertainty is comparable to the discrepancy between indicators seen in other z$\sim$1 works.  Our result that the shape of the SFRD as a function of stellar mass (and hence the mass range of galaxies dominating the SFRD) does not evolve between z$\sim$1 and z$\sim$0.1 is robust to the choice of indicator.

\noindent
\end{abstract}


\section{Introduction}
\label{sec:introduction}

Over the last decade or so, dozens of studies have measured the star formation rates of statistical samples of galaxies \citep[e.g.,][]{Lilly:1996la,Madau:1996ao,1999ApJ...519....1S,Hopkins:2006bv,Reddy:2009wc}, recently out to redshifts as high as z$\sim$6 and beyond \citep[e.g.,][]{Bouwens:2007tx}.  However, the majority of these are limited to the most actively star-forming and/or the most massive galaxies.  In order to construct a complete census of star-formation, it is necessary to probe to low enough star-formation rates (SFRs) to ensure that the star formation rate density (SFRD) has converged.  Star-forming galaxies may exhibit a relatively narrow sequence in star-formation rate as a function of stellar mass \citep{Noeske:2007tw}.  If this is the case, then the bulk of the high SFRs will be confined to the highest stellar mass objects.  However, since low mass galaxies are much more numerous than their higher mass counterparts, it is an open question in what mass objects the majority of the SFRD resides.  Even at z$\sim$1, when the Universe was 40\% of its current age, only the most massive galaxies (\ms$\gsim$10.5) are routinely explored by spectroscopic observations (e.g., \citealt{Cowie:2008ob,Damen:2009ph}).  The Gemini Deep Deep Survey (GDDS, \citealt{2004AJ....127.2455A}) recently managed to obtain spectroscopy of much lower stellar mass objects (\ms$\gsim$10) but this required large amounts ($\sim$30 hours) of integration on an 8-m telescope.  Lower masses can be investigated using photometric redshifts, but these either lead to larger uncertainties on the SFR for each using SED-fitting, since the redshift, SFR and dust must be fitted simultaneously \citep[e.g.,][]{Mobasher:2009qp}; or require statistical stacking of many objects using another SFR indicator such as radio continuum \citep{Dunne:2009bg,Pannella:2009uj}. The drawback with the stacking technique is that systematic errors in the stacked sample may bias the results, and only the properties of an average galaxy can be studied, not the variation between galaxies.

The combination of SFR and stellar mass data from surveys conducted at multiple epochs will provide important information on the evolution of star-formation and the build-up of stellar mass. For example, many recent studies \citep[][and others]{Noeske:2007tw,Cowie:2008ob,Dunne:2009bg,Pannella:2009uj,Damen:2009ph} are finding evidence for the `cosmic downsizing' of \citet{Cowie:1996xw} in which the dominant contribution to the SFRD proceeds to progressively less massive galaxies as the Universe ages.  However, such results have been based on direct observations of only the most massive galaxies at higher redshifts.  Thus, it is an open question whether low mass galaxies have always had similar SFRs at all times (and thus `downsizing' would simply be described by the decrease in the SFRD of high mass galaxies with cosmic time), or whether low mass galaxies show some different evolutionary behaviour.  In the latter case, it is conceivable that, at some point in the past, low mass galaxies may have dominated the universal SFRD.  Thus, a useful way to directly explore these scenarios is by looking at the SFRD as a function of stellar mass, down to low stellar masses (\ms$\sim$9),  and how this evolves with redshift. 
 
Ideally, one would like to obtain a stellar mass-selected survey, complete to low SFRs and repeat this at multiple epochs/redshifts.  To this end, we have designed the Redshift One LDSS3 Emission line Survey (ROLES), using guaranteed time on the LDSS3 spectrograph as part of the LDSS3 instrument project.  \citet{Davies:2009nx} (hereafter Paper I) used this survey of low mass galaxies at z$\sim$1 (a subsample of data from the current paper) plus higher mass galaxies from GDDS \citep[][hereafter J05]{Juneau:2005ft} and compared this with a local measurement from the Sloan Digital Sky Survey (SDSS).  In Paper 1, we found a suggestion that the \oii~luminosity density/SFRD at z$\sim$1 began to turnover below a mass of around \ms$\sim$9.5, and thus the bulk of the SFRD is in higher mass objects. Paper 1 presented initial results from a subsample of three masks in one of our pointings.  Here we present all the data from the same redshift bin for both our fields (three different pointing positions), resulting in five times the number of z$\sim$1 galaxies due to the greater areal coverage and higher spectroscopic completeness.  We also employ a new empirical correction (\citealt{Gilbank:2009nx}, hereafter \g09) to accurately correct the mass-dependence of \oii-SFRs; SFRs from SED-fitting; new higher mass samples from other spectroscopic surveys; and an improved local comparison sample built from the SDSS.

This paper presents the survey design and method and results using the complete dataset from the $0.89<z<1.15$ component of the survey. The layout of this paper is as follows.  \S2  describes the survey design, observations, and data reduction. \S3 describes the analysis of the catalogues so produced and comparison samples from other z$\sim$1 surveys.  \S4 shows and discusses the results from the \oii~luminosity function and star formation rate density, and \S5 concludes.  All magnitudes are quoted on the AB system unless otherwise stated, and we assume a cosmology $(h,\Omega_M, \Omega_\lambda)=(0.7,0.3,0.7)$. Throughout, we convert all quantities to those using a \citet{Baldry:2003ue} (hereafter BG03) universal initial mass function (IMF).

\section{Method}

\subsection{Survey Design}
ROLES (the Redshift One LDSS3 Emission line Survey) is designed to probe the SFRD in galaxies with stellar masses much lower than previously studied at redshifts of order unity.  In order to do this efficiently, we adopt a novel survey strategy.  We utilise fields with very deep $K$-band photometry and photometric redshifts in order to pre-select galaxies which are most likely low-stellar mass systems at z$\sim$1.  These galaxies are then followed up with multi-object optical spectroscopy in order to obtain both the redshift and star-formation rate.  To increase efficiency even further, we use a set of custom band-limiting filters to restrict the wavelength range (and hence the corresponding redshift range for a given rest-frame wavelength) of our spectra and provide us with a high sampling density.  We select $K$-faint\footnote{all our magnitudes are on the AB system unless otherwise noted.  $K_{\rm AB}=K_{\rm Vega}+1.87$} ($22.5 < K \le 24.0$) targets.  Since securing absorption line redshifts for galaxies this faint is extremely challenging (c.f. the brighter sample of \citealt{2004AJ....127.2455A}, GDDS) and would require prohibitive amounts of telescope time, we choose to use relatively modest exposure times ($\sim$ 4 hours on a 6.5-m telescope) and specifically target the \oii~emission line within our redshift window.  With this approach, we will not obtain redshifts for galaxies without emission lines (to our flux limit), but these do not contribute significantly to the SFRD.  In this way, we obtain a mass-selected sample, complete to a given SFR limit (to the extent that \oii~traces the SFR).

\begin{figure*}
	{\centering
	\includegraphics[width=140mm,angle=0]{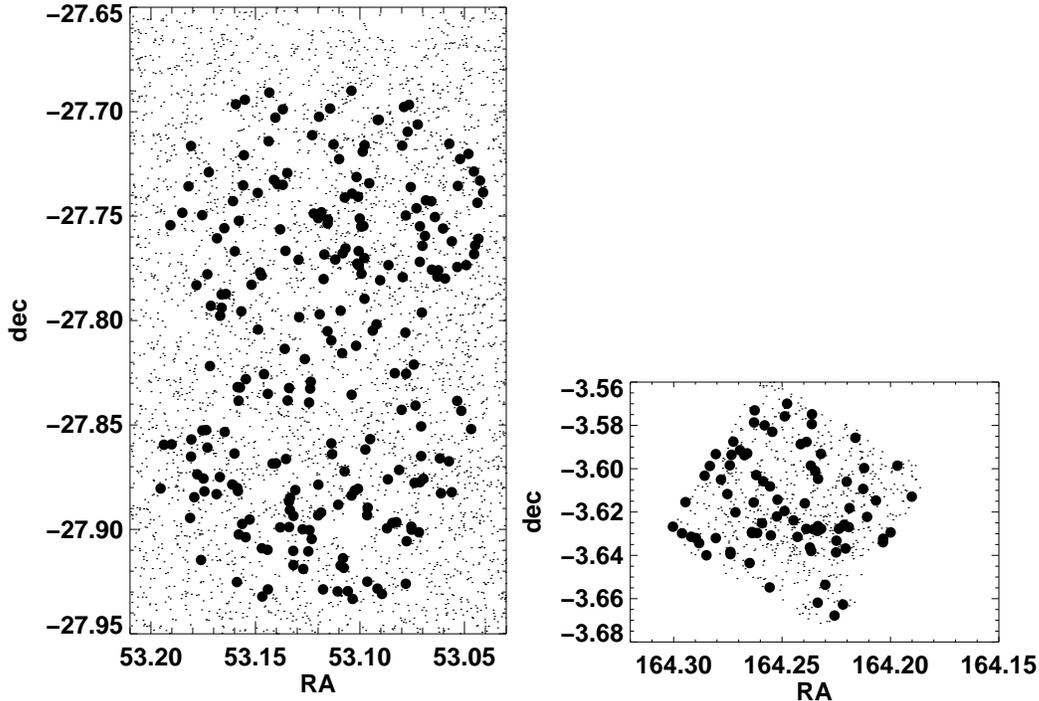}
	\caption{Layout of the ROLES fields: CDFS (left) and FIRES (right) on the same scale.  CDFS comprises two pointing positions centred on 03$^h$32$^m$27.6$^s$, -27$^d$45$^m$00$^s$ and 03$^h$32$^m$28.8$^s$, -27$^d$52$^m$12$^s$ of roughly 8 arcmin diameter, set by the field of view of LDSS3.  FIRES is a single pointing of 5.5$\times$5.3 arcmin, set by the limit of the deep photometry available.  Points indicate galaxies with 22.5$<K<24.0$ and filled circles show our detections of z$\sim$1 emission line galaxies.}
	 \label{fig:fields}
}
\end{figure*}

Fields with both the depth and wealth of multi-colour imaging we require for this survey are currently scarce.  For ROLES, we use photometric catalogues and photometric redshifts from two imaging surveys in two different fields to select targets for follow-up spectroscopy: the Great Observatories Origins Deep Survey (GOODS) region of the {\it Chandra} Deep Field South (CDFS, e.g., \citealt{Wuyts:2008lq}), and the Faint Infra-Red Extragalactic  Survey (FIRES, e.g., \citealt{Forster-Schreiber:2006ih}). We note that FIRES actually comprises two different fields: the Hubble Deep Field South (HDF-S) and a region containing the z$=$0.83 cluster MS1054-03.  In ROLES we have only followed up the MS1054-03 area (due to its significantly wider area than the HDF-S), but hereafter we refer to the MS1054-03 field as ``FIRES". The  layout of our fields is shown in Fig.~\ref{fig:fields}.

\subsection{Sample selection \& Spectroscopic Observations}
\begin{figure}
	{\centering
	\includegraphics[width=60mm,angle=90]{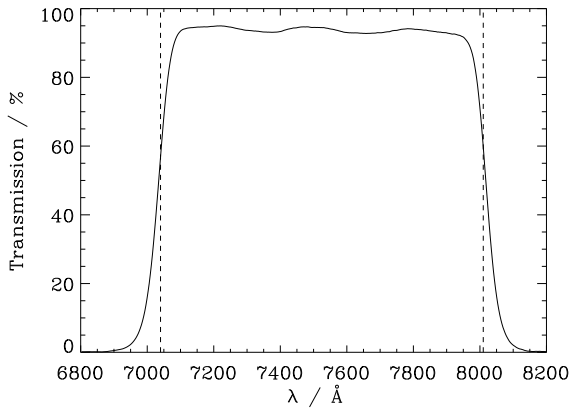}
	\caption{The transmission curve of the KG750
	filter.  The dashed lines at the half-maxima denote the minimum
	and maximum wavelengths used, 7040\AA--8010\AA.}
	 \label{fig:filter}
}
\end{figure}

For our LDSS3 spectroscopy, we obtained a set of four custom-band
limiting filters to observe galaxies in well-defined redshift bins.
In this work, we only consider the KG750 filter which covers
approximately $(750\pm50)$ nm.  Observations using the KG650 [(650$\pm$50) nm] are ongoing and will be presented in future
work.  The transmission curve of the KG750 filter is shown in
Fig.~\ref{fig:filter}.  Half-maxima occur at 7040\AA~and 8010\AA~and
 these are adopted as the wavelength limits.  In targeting the \oii~line
at 3727\AA, these limits correspond to a redshift range of $0.889 < z
\le 1.149$.

LDSS3 can perform multi-object spectroscopy (MOS) in nod-and-shuffle
(N\&S) mode \citep{2001PASP..113..197G}, which allows extremely
accurate sky-subtraction thus allowing lower flux limits to be reached 
than would otherwise be achievable with traditional MOS.  This is
particularly important in the wavelength window of interest here, which is populated by numerous bright sky lines.  The LDSS3 medium red grating (300 lines/mm was used, giving an average dispersion
of $\sim$ 2.7\AA/pixel) and 0.8\arcsec~wide slits which, with the
plate scale of 0.189\arcsec/pixel, gives a resolution of 11.4\AA~FWHM.
The N\&S observations utilised 3\arcsec long slits, observing each
target galaxy for 60s at one end of the slit and then nodding the
telescope 1.2\arcsec and observing for another 60s at a second
position within the slit.  These two positions are referred to as A and B
for convenience.  By nodding within the slit, the galaxy is observed 
for the total time the detector shutter is open.  Charge is shuffled
along the detector between the A and B observations by 16 pixels (corresponding to the slit width of 3\arcsec), resulting in two exposures for each galaxy, stored in different locations on
the CCD.  Because the telescope was nodded between exposures, the B
exposure contains observations of the sky at the location of the
galaxy in the A exposure and vice-versa.  Thus sky-subtraction can be
achieved simply by subtracting the A observation from the B.  This
process is described in more detail in the next section.

With the above parameters, we can place an average of almost 200 objects in each mask over the $\sim$8.2 arcmin diameter circular field of view of LDSS3.  To select targets, galaxies were prioritised by assigning a weight to them based on their $K$-band magnitude, photometric redshift and its error.  Galaxies in the range $22.5 < K \le 24.0$ were primarily targeted.  Photometric redshifts and confidence intervals kindly provided by B. Mobasher and T. Dahlen (CDFS) and from \citet[][FIRES]{Forster-Schreiber:2006ih} were used to give higher weighting to those galaxies whose error bars placed them within (or overlapping with) the redshift window $0.889 < z \le 1.149$.  Once weights were assigned to each galaxy, slits were randomly allocated to objects, sorted by priority.  Due to geometrical constraints, once all the highest priority targets were assigned, the mask was filled with lower priority targets (typically more uncertain photometric redshifts or $K$ magnitudes outside our main sample).  The total weight for all targets allocated in the mask was then calculated and a Monte-Carlo technique used to design many realisations, keeping the version with the highest weight.  Multiple masks at the same pointing position were then designed, giving highest priority to those galaxies not already observed in a previous mask (but repeats of some slits occur and are useful for checking purposes, discussed later).  The exact details of the object prioritising are unimportant as a) the completeness is calculated {\it a posteriori} in \S\ref{sec:completeness}, and b) eventually enough targets are observed to ensure high completeness for all galaxies within this $K$ magnitude range after $\sim$6 masks.  The main reason for this prioritisation is to ensure a relatively high completeness after only a few ($\sim$3) masks have been observed. 

We observed each mask for typically 20 N\&S cycles of 60s exposures at each of the A and B positions, resulting in an exposure time of 40 minutes for each image.  We typically took 6 exposures resulting in a total integration time of four hours.  A log of the observations is given in Table \ref{table:masks}.  The seeing ranged from 0.6\arcsec~to 1.0\arcsec~and was typically around 0.8\arcsec.  It should be noted that the few frames with significantly shorter exposure times are lower priority masks and so the overall loss in completeness if these masks are slightly shallower is minimal. The final numbers of slits we are able to place on ROLES' targets (i.e., matching our $K$-band selection criterion and photo-z weighting) total 1849 for CDFS and 533 for FIRES. These totals refer to the number of spectra we might expect to obtain for unique objects (i.e., repeated slits are only counted once)  after discounting `filler targets' from other programmes and the few objects ($\lsim$5 per mask) which resulted in corrupt spectra due to poorly-cut slits, slit collision or contamination by scattered light, etc.

\begin{table}
	\caption{Summary of the observations.  The two pointing positions for CDFS are denoted CDFS.1 and CDFS.2 (northern and southern pointing, respectively).  Total exposure times include only data of acceptable quality used in the analysis.  Results based on masks indicated with ``$^*$" were presented in paper 1, but these have been re-reduced with the present method for uniformity. 	}
	\label{table:masks}
	\begin{tabular}{lll}
		\hline
		Mask & Total exposure & Field \\
		           & time /$ks$          & \\
		\hline
mask1   &    31.2  &	FIRES    \\
mask2   &    24.0  &     FIRES	 \\
mask4$^*$   &    21.6  &     CDFS.1	 \\
mask5$^*$  &    14.4  &     CDFS.1	 \\
mask6$^*$  &    12.0  &     CDFS.1	 \\
mask7   &    19.2  &     CDFS.2	 \\
mask8   &    19.2  &     CDFS.2	 \\
mask9   &    16.8  &     CDFS.2	 \\
mask10  &     9.6  &     CDFS.1	 \\
mask11  &     13.2 &     CDFS.1	 \\
mask12  &     15.6 &     CDFS.1	 \\
mask13  &     9.6  &     FIRES	 \\
mask14  &     16.8 &     FIRES	 \\
mask15  &     13.2 &     FIRES	 \\
mask16  &     10.8 &     FIRES	 \\
mask21  &     15.6 &     CDFS.2	 \\
mask22  &     18.0 &     CDFS.2	 \\
mask23  &     15.6 &     CDFS.2	 \\
		\hline \\

	\end{tabular}
\end{table}

\subsection{Spectroscopic data reduction}
\label{sec:data_reduction}

\subsubsection{Pre-processing}

\begin{figure*}
	{\centering
		\includegraphics[width=140mm]{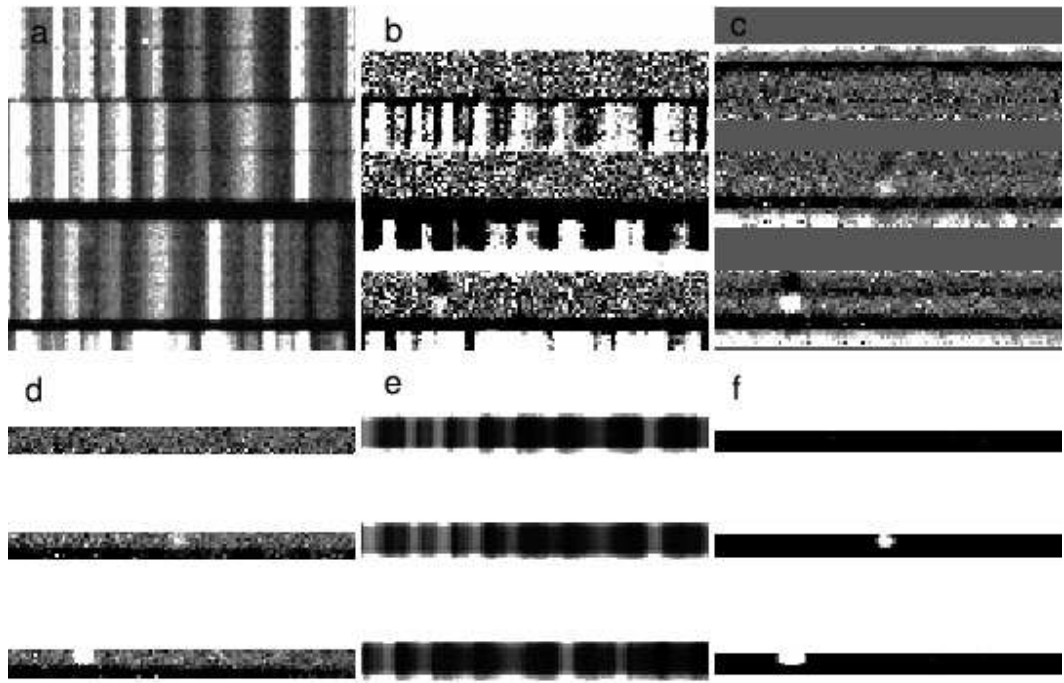}
	\caption{Examples of the various steps in the data reduction, showing a region containing three representative slits.  	The six panels each show a region of 100 pixels (270\AA) in the spectral (horizontal) direction by 100 pixels in the spatial (vertical) direction.
		Panel a): the initial individual frame.  Panel b): the same frame after sky-subtraction (by shifting a copy of the image by 16 pixels and subtracting) 
	Panel c): after median combination of several exposures with cosmic ray rejection.
	Panel d):  coaddition of the positive and negative (A$^\prime$-B$^\prime$) spectra after subtracting the two different nod positions  Panel e): the noise frame calculated as described in the text.  The highest levels of noise correspond to regions underneath bright sky lines. Panel f): The final signal-to-noise map.  Black corresponds to a level of 3.0$\sigma$ and white to 7.0$\sigma$.  Two peaks corresponding to emission lines clearly stand out.  Tracing them back to earlier stages of the reduction, the line in the lowest slit can be seen in an individual exposure (panel b), but the line in the central slit only becomes vsisible after stacking several frames (panel c).  The dipole signature of a genuine emission line is clearly visible in the two individual nod position spectra (A$^\prime$ and B$^\prime$).
	}
	 \label{fig:frames}
}
\end{figure*}

In addition to standard spectral CCD data pre-processing (overscan subtraction, bad pixel masking etc.) a number of non-standard steps are necessary for our data, which we detail here.  We used several routines from the Carnegie Observatories System for Multi-Object Spectroscopy 2 ({\sc COSMOS2}\footnote{see http://users.ociw.edu/oemler/COSMOS2/COSMOS2.html}), but the majority were applied using custom written routines in IDL.  Examples of three representative slits from one mask are shown at various stages of the reduction in Fig.~\ref{fig:frames}.  In all cases the same portion of the frame is shown, which is about 100 pixels ($\sim$270\AA) long in the spectral direction.  Panel ``a" shows a single, raw exposure prior to sky subtraction.  

Firstly the {\sc COSMOS2} {\sc stitch} routine is used to combine the images from the two separate LDSS3 amplifiers into a single frame and convert the counts in the image into electrons. Overscan subtraction and bad pixel masking were then performed (the bad pixels being identified by significantly deviant pixels seen in the ratio of two dispersed sky flats).  

\subsubsection{Charge trap and bad pixel masking}

The repeated charge shuffling involved with N\&S observations can cause artefacts which appear as short streaks along certain columns of the CCD.  This is due to localised defects (`charge traps') in the CCD which cause the charge to become smeared as it passes over them.  Charge traps are identified by taking dark frames using the same N\&S strategy as for the science data.  This leads to a frame containing columns of bright pixels along the shuffle direction due to charge trap defects.  These are identified by median smoothing the image using a 1$\times$3 pixel boxcar (so that the long axis is perpendicular to the shuffle direction) and taking the ratio of the original N\&S dark to the median-smoothed version.  Charge traps show up as significantly deviant pixels and are flagged and added to the original bad pixel mask.  After using this charge trap mask only a very small number of obvious charge traps still remain and we discuss how these are dealt with in \S\ref{sec:line_detection}.
The same charge trap mask is used for data from all observing runs, since the pattern appeared stable and the few residual charge traps are removed by hand, as described below. 

\subsubsection{Registration and wavelength calibration}

The next step is to calculate the transformation between each individual exposure, to register all exposures of the same mask to a common frame.  There are two consequences leading to offsets caused by flexure within the telescope and/or instrument (or not perfectly reproducible positioning of the slit mask within the instrument).  The first means that the images of the slits may not fall on the same pixels each time.  This is especially the case when data were taken on different observing runs.  To deal with this, we use the mapping routines in {\sc COSMOS} ({\sc map-spectra}, {\sc adjust-map}, etc.) to accurately identify the locations of a set of reference sky lines in every single exposure.  By taking the first frame as a reference and requiring the positions of the sky lines to match in each subsequent exposure, we can calculate the linear shift and rotation (and occasionally a small scale change) to register the images\footnote{Since LDSS3 is oversampled, we use nearest-neighbour resampling for all transformations, to avoid correlating the noise between neighbouring pixels and hence compromising the noise estimate in each pixel.}.  The second possible offset involves the movement of the galaxy within the slit and will be dealt with later.  Up to this point, we have calculated how to align the images of the sky spectra. 

{\sc adjust-map} in {\sc COSMOS} does a good job of accurately centroiding on the sky lines in the dispersion direction, however significant offsets are seen from the centre of the slit in the spatial direction.  This may be because the routine is not optimised to work with N\&S data.  Since accurate slit location is essential (as described later) we use our own slit tracing routine to overcome this.  Briefly, this code takes many 1D profiles through each spectrum in the spatial direction and cross-correlates this against a model of a N\&S spectrum (approximately two top-hats of width 16 pixels, the slit-width/shuffle distance, side-by-side) to find the optimum offset.  In the tracing and rectification, slit curvature and tilt are neglected.  Since the spectra are short (370 pixels), the change in y coordinate over this length is typically one pixel or less, so spectra are extracted in rectangular boxes aligned to the CCD axes, for simplicity.

Before applying this transformation and resampling the spectra, it is necessary to perform the sky subtraction as described below.  If this is not done, a small rotation can result in strong sky-subtraction residuals due to the sharp edges of bright sky lines being under/over-subtracted as they pass into a neighbouring pixel.

As a by-product of calculating this transformation, maps are constructed to convert global x and y CCD coordinates to wavelength and slit spatial position (specifically slit centre).  These maps will later be used to identify the wavelength of a detected emission line, transforming from its CCD coordinates.

\subsubsection{Sky-subtraction}

For each individual exposure (see Fig.~\ref{fig:frames}, panel a), N\&S sky subtraction is performed by shifting the image by the shuffle
distance (16 pixels) in the spatial direction and subtracting it from
itself (Fig.~\ref{fig:frames}, panel b).  If we consider A-B, in the notation introduced in the
previous section, every slit will now contain a positive image of
galaxy spectrum A$^\prime$ (where the prime is used to denote a
sky-subtracted spectrum) and a negative image of spectrum B$^\prime$
of the same galaxy, both with the sky removed.  

Once the sky subtraction has been performed the spectra are resampled to  a common frame, as described above.  Using the boundaries identified by the slit-tracing program, all pixels outside the spectra are flagged as bad and their values omitted from further analysis.  This is crucial to avoid contamination of the edges of regions of good data with the bright sky spectra at the edge of the slit.  This also ensures that the correct exposure time is propagated (by tracking the number of good exposures in each pixel after the next alignment step).  The next step is to align the positions of the galaxies within the slits.  Although the positions of the slits have been registered, the position of the galaxy can shift by up to a couple of pixels due to alignment drift of the instrument. To register the positions of the object spectra, several bright nebular emission lines in the galaxy spectra are identified in each mask.  The centroids of these lines are measured in each individual exposure and offsets calculated.\footnote{Since only a small number of objects per mask ($\lsim$6) possessed emission lines bright enough to be found visually in an individual exposure, we were unable to fit for any transformation other than a small shift.  It is possible that the galaxy positions may be rotated between exposures, for example.  To assess the accuracy of this transformation, we calculated shifts from three bright lines and looked at residual shifts in other bright lines in the same exposure.  Typically these were centred to better than 2 pixels.}  For all masks, the shifts are found to only be significant in the y (spatial) direction (confirming the accuracy of the wavelength calibration) and the spectra are then resampled to the nearest integer pixel shift.

Next,  all the individual exposures are median-combined with 5$\sigma$ outlier-rejection (where $\sigma$ is the noise estimate, described below) to remove cosmic rays (Fig.~\ref{fig:frames}, panel c). (Experimentation showed that median combination gave a slightly cleaner co-addition than using the mean.) In order to be left with a positive, summed image of the galaxies, the sky-subtracted spectra can simply be shifted by their separation, 6 pixels (which is equal to the nod distance, 1.2\arcsec), and subtracted, A$^\prime$-B$^\prime$ (Fig.~\ref{fig:frames}, panel d). Notice that in Fig.~\ref{fig:frames} panel d, pixels outside the region of the combined science spectra are set to bad values (appearing blank in the plot).  Prior to object-finding, the extracted 2D spectra are shrunk by one pixel in the spatial direction (i.e., above and below) on either side of the slit.  The nearest neighbour resampling technique means that all transformations are accurate at the one pixel level, and so this prevents rounding error.  However, it can be seen that in the examples shown, bright sky lines are bleeding into the lower portion of the slit (appearing black in the image, since they have been subtracted in the previous step). Shrinking the extraction window helps to reduce the impact of such artefacts. However, since the spatial extent of the emission line is very close to the width of this window, care must be taken to compensate for flux which may fall outside this window (the top edge in this case), discussed in \S\ref{sec:flux}. 

This then leaves us with the final, reduced 2D spectra\footnote{One step which has been intentionally omitted is flatfielding. Tests running the data through our whole analysis show that attempting to flatfield makes negligible difference (the N\&S methodology will also naturally average over some pixel-to-pixel sensitivity variations) and possibly marginally increases the noise.  The main limitation to the accuracy of the flatfielding is the accuracy with which the flatfields can be registered to the science data.  The flatfields cannot be registered in the same manner as the science data since the latter rely on the 2D position of night sky lines.  The flatfields may only be wavelength calibrated from arc lamps, taken periodically through the night, and by the positions of the slit edges to locate the spatial offset.  This can lead to small systematic offsets between the flatfield and the science data which may lead to increased noise when the (offset) flat is applied.} from which we will make measurements such as line flux.  However, in order to optimally detect emission lines in the first place, we will further process the images to produce optimally filtered versions.  We also require an estimate of the noise in each pixel of our detection images to assess the significance of our detections, which are described below.

\subsubsection{Noise estimation}

By following a similar procedure to the sky-subtraction just described (i.e. shifting a copy of the image by the shuffle distance), but {\it adding} instead of {\it subtracting} the frames, an image is produced which contains a spectrum of the sky instead of the galaxy at the positions of the galaxies (A and B) in each slit.   
By again median-combining the individual exposures a frame is created with the best estimate of the value of the sky at each pixel.  This is used to estimate the noise, $N_{indiv}$, in each pixel of each nod position in an individual exposure.  The noise in the $i,j^{th}$ pixel is given by 

\begin{equation}
	N_{indiv, ij} = \sqrt{|<sky>|_{ ij} + 2R^2},
	\label{eqn:dr1}
\end{equation}
where $<sky>$ is the median-combined sky frame, $R$ is the read noise of the detector (3.4 electrons), and the factor of 2 accounts for the fact that two readouts have been combined in the sky-subtraction/addition by using the two shuffle positions. 

After stacking $n_{frames}$ exposures, the noise in the $i,j^{th}$ pixel, $N_{com, ij}$, becomes

\begin{equation}
	N_{com, ij} = N_{indiv, ij}/\sqrt{n_{frames}}.
\end{equation}

Once the two nod positions in the combined frame produce an image containing the sum of the objects, the noise in the $i,j^{th}$ pixel of the final combined science image is then

\begin{equation}
	N_{ij} = \sqrt{N_{com_{A'},ij}^2+N_{com_{B'},ij}^2},
	\label{eqn:noise1}
\end{equation}
which is just the quadrature sum of the A$'$ and B$'$ position noise values (Fig.~\ref{fig:frames}, panel e).

\subsection{Line Detection}
\label{sec:line_detection}

\subsubsection{Optimal filtering}

We wish to search the 2D spectra objectively for significant features consistent with
emission lines from the target galaxies (see also \citealt{Glazebrook:2004yf} for a similar application).  Since the optimal detection
kernel is the profile of the line itself, we start by finding several
obvious bright lines, easily visible by eye, and measuring their
profiles.  These are typically well-approximated by
elliptical Gaussians with FWHM of 5 pixels in the spectral direction
and 3.5 pixels in the spatial direction.  The kernel is normalised by
setting its total to unity to conserve flux.  

The science images are convolved with this kernel, producing a signal frame, $S$ (Fig.~\ref{fig:frames}, panel f).  This convolution changes the properties of the noise in each pixel from $N_{ij}$ to $N_{conv, ij}$ where

\begin{equation}
	N_{conv,ij} = \sqrt{N_{ij}^2 \otimes k^2},
\end{equation}
where $k$ is the emission line-shaped kernel described above.
$N_{conv,ij}$ is thus the quadrature sum of the contributions of each
pixel in the kernel to the noise in the $i,j^{th}$ pixel. 

Now, for the $i,j^{th}$ pixel, the significance, $\sigma$,  is given by 
\begin{equation}
	\sigma_{ij}=\frac{S_{ij}-b_{ij}}{N_{conv, ij}}
	\label{eqn:dr2}
\end{equation}

\noindent
where $b_{ij}$ is the local background (continuum) estimated from the mean of all pixels in two 1D side-bands, each 40 pixels wide, ranging from $i+10$ to $i+50$ pixels and $i-10$ to $i-50$ (or whatever number of pixels from these regions falls on pixels of the science spectrum flagged as good).  It is appropriate to only measure the continuum in a 1D box (i.e. a line at the position of the $j^{th}$ pixel) in this way, since the image has already been convolved with a kernel in the spatial ($j$) direction and thus contains an estimate using values from multiple pixels in this direction.  This simple approach to continuum estimation has the drawback that it can be severely biased high when a bright emission line enters one of the sidebands.  To alleviate this problem, we perform one iteration, re-estimating the continuum in the same way after masking out all pixels above a high significance threshold (6$\sigma$) from the first pass.\footnote{This threshold was determined by experimentation. It was found that emission lines with an initial significance $<$6$\sigma$ had negligible impact on the estimate of their surrounding continuum, and thus this threshold is a conservative threshold for re-estimating the continuum level.}

The above is an exact expression considering sky noise and read noise.  However, in some cases we cannot neglect the contribution to the noise from the Poisson noise in the continuum (even for these faint objects).  We adopted the following approach as an operational way to consider the effects of continuum noise.  

The maximum contribution from the continuum is taken as the maximum within one pixel above or below the position of the candidate emission line:

\begin{equation}
b_{max}=max\{b_{ij-1}, b_{ij}, b_{ij+1}\}
\end{equation}
(to allow for the effect of rounding error in locating the position of the continuum).  Then, the noise in the kernel-smoothed image is estimated as

\begin{equation}
b_{noise} = \sqrt{b_{ij} \otimes k^2},
\end{equation}
where the $b_{ij}$ term here is from the Poisson noise squared.  The final significance estimate is then modified to 

\begin{equation}
\sigma_{ij}=\frac{S_{ij}-b_{max}}{\sqrt{ N_{conv, ij}^2 + b_{noise}^2  } }
\end{equation}
for sources with significant flux in the continuum.  Significant flux can be as little as $\sim$5-10 electrons, since other sources of noise, even bright sky lines, have been beaten down by repeated observations.

We calculate the significance of each pixel in the science spectra and locate the most significant peaks.  Such a significance map is shown in Fig.~\ref{fig:frames} panel f.   Significant pixels are grouped together using a standard n-connected pixels algorithm (joining all neighbouring pixels above the minimum significance threshold into a single group), identifying the peak by the single most significant pixel.  In addition, detections are rejected where the peak pixel lies at the edge of the spectral extraction box, as these are always due to edge effects from bright sky lines.

\subsubsection{Catalogue cleaning}

The majority of artefacts present in the emission line catalogue come from one of two sources.  There are: bright sky lines leaking into the edge of the slit extraction box, and broad `lines' caused by weak continuum emission which is not well sampled by our simple sliding box continuum filter.  These all needed to be flagged interactively.

\begin{figure*}
	{\centering
	\includegraphics[width=160mm,angle=0]{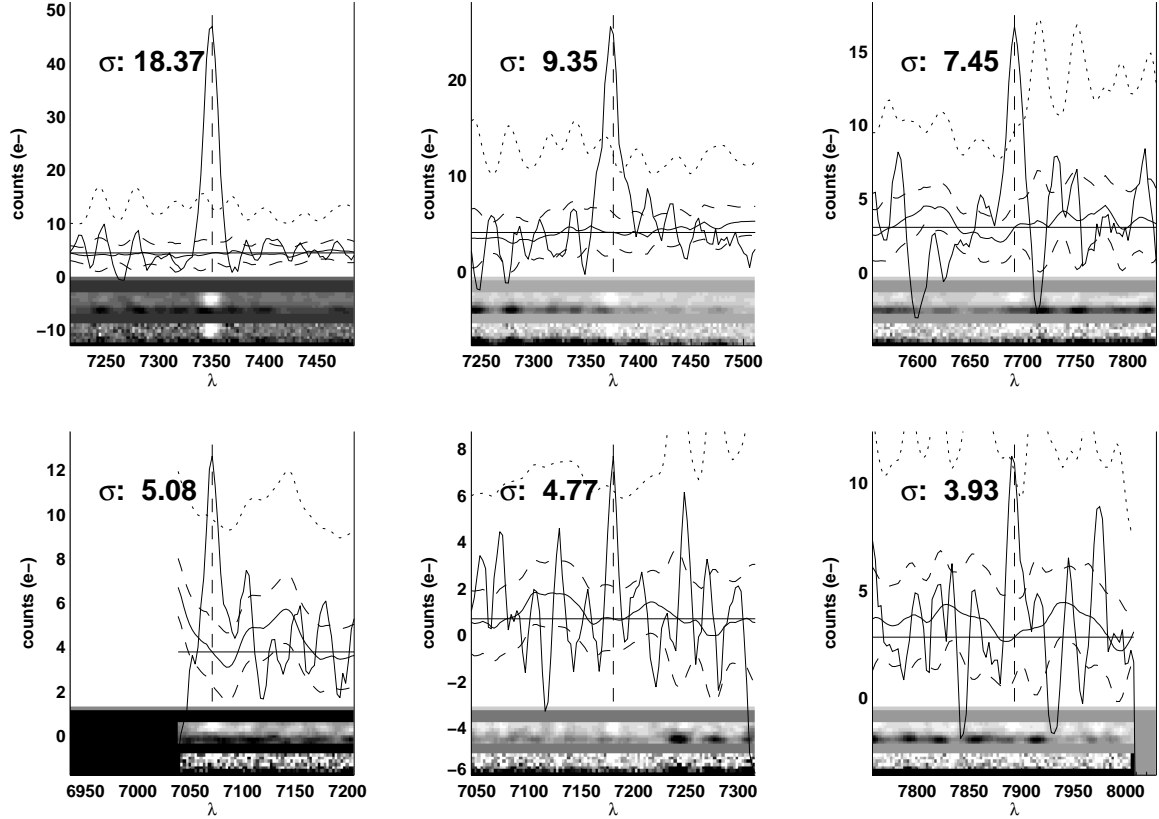}
	\caption{Examples of 1- and 2-D spectra, ranked in order of decreasing significance. Each panel shows the smoothed 1D spectrum (thick black line), centred on the detection, as the main plot. The solid  line bounded by two dashed lines shows the estimate of the continuum and the 1$\sigma$ noise limits around this, respectively, and the dotted line indicates the nominal 4$\sigma$ threshold.  The plot is annotated with the significance of the detection. The two insets below show the smoothed (upper inset) and unsmoothed (lower inset) 2D spectrum.  The wavelength scale is shown below.  
	 \label{fig:oned}
}}
\end{figure*}

Fig.~\ref{fig:oned} shows examples of convolved and un-convolved 2D and 1D spectra for
a range of line significances.  Such plots are visually inspected, sorted by
significance, for all objects.  Only detections which are due to obvious artefacts (edge effects from sky lines, missed charge traps, scattered light from bright neighbouring spectra, underestimated continuum) are rejected.  We make no additional cuts  on a subjective basis, such as line shape, for example.  A reasonably clear cut-off in the reliability of lines is seen at $\sim4\sigma$ in our units using this technique, but this will be re-assessed quantitatively in the next section.  Above this limit, the
shapes of the lines in all the plots appeared broadly consistent with
each other. After cleaning the catalogue, we are left with emission line detections for 246 unique objects in CDFS and 119 in FIRES, from the 1849 and 533 unique objects (CDFS and FIRES, respectively) on which we initially placed slits. There are several tests which can be performed to assess the reliability of our emission line catalogue.

\subsubsection{Tests of emission line catalogue}
\label{sec:tests}

The most basic quality check is to compare the properties of emission lines detected in the catalogue with repeat observations of the same galaxy observed in a different mask.  In this comparison we select galaxies from our emission line catalogue and identify which of these have been observed in more than one mask.  

\begin{figure}
{\centering
\includegraphics[width=90mm]{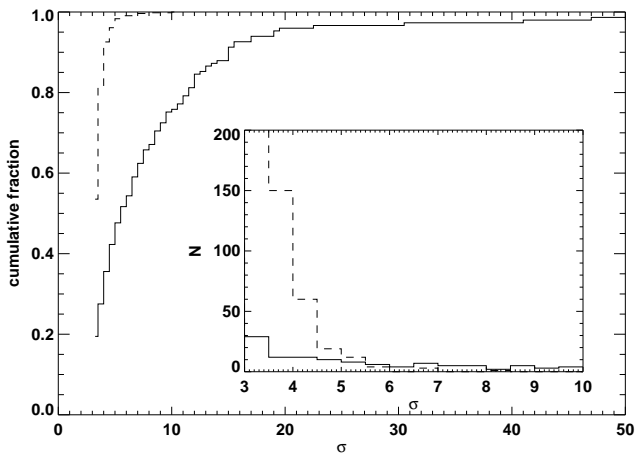}
\caption{Reproducibility of lines from repeated observations of the same galaxies in mutliple masks.  Solid lines show those lines successfully reproduced in multiple masks, dashed lines show objects which weren't successfully independently found on every repeated mask.  See text for further details.  Main panel shows the cumulative fraction of such detections, indicating that 95\% of `spurious' identifications occur at significance $<$4.0$\sigma$.  Inset shows a histogram of the number of detections, zoomed in to the low significance end of the plot.
\label{fig:repeats}
}
}
\end{figure}

There are 44 such repeat slits between these masks (observed a total of 100 times - some objects are observed on several different masks).  Each detection is tested in turn to see if it has been reproduced in a repeat observation of another mask.  Each detection is considered individually (some z$\sim$0.5 galaxies have as many as three genuine emission lines in each slit).  If it is independently detected on  every mask, we consider this successfully reproduced.  If it is not found on {\it every} mask, it is added to a list of non-reproduced objects.  We note that this is conservative, as an object independently discovered in two out of three masks would be classed as spurious in this scheme.  Also, there may be some overcounting in this method as it is possible that if a galaxy is observed in two masks and both produced discordant detections, both would be counted as spurious when in fact the more significant line may be genuine for the following reason.  Due to the slightly different flux limits of the different masks, it is possible that a line observed in one mask may be just below the detection threshold in a shallower mask.  We make no attempt here to correct for the different depths, and thus a line overlapping with a slightly shallower mask may be undetected in the shallower mask and thus incorrectly flagged as spurious.  Regardless, using these criteria, the objects successfully recovered are plotted in Fig.~\ref{fig:repeats} (solid line) and those not recovered (dashed line). The cumulative fraction (main panel) shows that 95\% of all false detections\footnote{The line detection algorithm stores all detections down to 3.0$\sigma$.} occur below 4.0$\sigma$.  We reiterate that this is likely an upper limit, for the reasons just discussed.

This internal comparison also allows an estimate of the accuracy of the wavelength calibration.  The difference in wavelength between lines successfully recovered is well described by a Gaussian of width $\sigma$=1.4\AA, which is $\sim$0.5 CCD pixels.  This corresponds to an observed frame velocity difference of $\sim$100 \kms~for ROLES galaxies or $\sim$50 \kms~rest frame at z$\sim$1.

\begin{figure}
{\centering
\includegraphics[width=90mm]{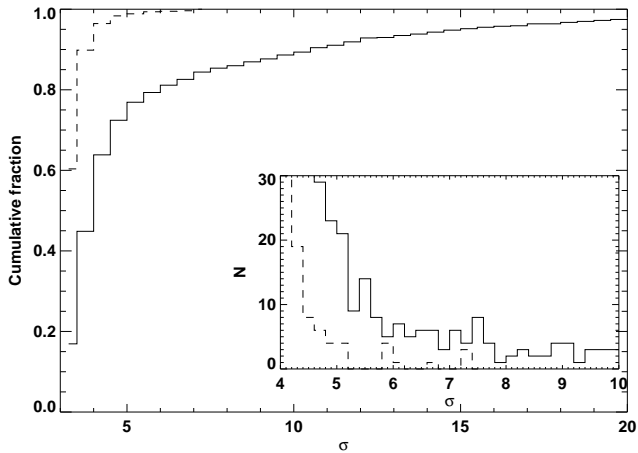}
\caption{Using the combination of multiple lines within each slit to assess the genuine/spurious nature of detections.  Solid lines show slits where the combination of lines yield a self-consistent redshift, dashed lines show those which do not.  Main panel shows the cumulative fraction of such detections, indicating that 90\% of `spurious' identifications identified in this way occur at significance $<$4.0$\sigma$.  Inset shows a histogram of the number of detections, zoomed in to the low significance end of the plot.  See text for details.
\label{fig:mspurious}
}
}
\end{figure}

A second internal test of the catalogue is possible for those galaxies with multiple emission lines.  We can attempt to measure redshifts from multiple lines and check that the wavelengths of the lines detected are consistent with a single redshift for the object.  This is done by considering all possible combinations of common galaxy emission lines, compared with the observed wavelengths of our detected lines. Plots of these test redshifts for each galaxy are visually inspected and the most likely combination assessed.  In almost all cases there is either one or zero match.  For a small number of cases, implausible combinations of lines are rejected (such as \nii, \sii, but no H$\alpha$).  Other than this, no astrophysical considerations of likely line ratios are used to make the decision.  Where no plausible combination is found, the most significant line is retained as a single line detection and all lower significance lines in that slit are rejected.  This may not be a completely representative number for the whole survey as some slits contribute several lines to the spurious count by this measure.  
To reduce the impact of the few slits with many false positives, we only count a maximum of one false positive per slit.  This is likely a better estimate of the true rate in our survey, since most slits have at most one detection. In a similar manner to Fig.~\ref{fig:repeats},  Fig.~\ref{fig:mspurious} shows the likely spurious detection rate arising from non-physical combinations of lines.  Note that 90\% of spurious lines by this measure occur below a significance of 4.0$\sigma$. 

75 objects result in plausible multiple-line redshifts. Of these, 61 belong to z$<<$1 galaxies (mostly z$\sim$0.5); 14 show \oii~at z$\sim$1 plus at least one other emission line.  

\begin{figure}
{\centering
\includegraphics[width=100mm,clip=t]{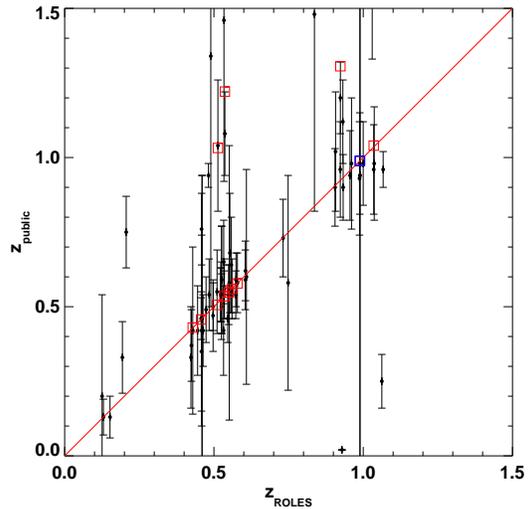}
\caption{Comparison of secure spectroscopic redshifts from ROLES galaxies possessing multiple emission lines with independent redshift measurements from the literature.  Filled circles with error bars denote photometric redshifts from the FIREWORKS and FIRES catalogues; red squares denote secure spectroscopic redshifts from the sources discussed in the text; blue squares show lower confidence ($\le0.5$) public spectroscopic redshifts. The vast majority of public redshifts agree well with our multiple line redshifts, and we discuss in the text the reasons for the small number of outliers (which are likely more common in this comparison subsample using multiple lines than for ROLES as a whole, where the typical object exhibits a single line).  
\label{fig:zs_sec_zp}
}
}
\end{figure}

As a further test for these multiple line objects (secure redshifts) we can compare the redshifts from our data with others available from the literature.  Where possible, public spectroscopic redshifts are used.   For CDFS we use the compilation from \citet{Wuyts:2008lq} and only consider secure redshifts (their quality flag 1.0).  This CDFS sample comes from a variety of surveys with very different selection criteria and thus caution must be used when considering how representative of the whole of ROLES are any statistics derived from a comparison with these redshifts.  For the FIRES field, we use spectroscopy from \citet{Tran:2007hx} (kindly provided by K.-V. Tran) primarily designed to study the z$=$0.83 cluster, MS1054-03, below the redshift limit of ROLES. Again, due to the very different selection criteria from ROLES' galaxies, care should be exercised in drawing comparisons.   A plot of the ROLES redshifts against these others is shown in Fig.~\ref{fig:zs_sec_zp}.  Points with error bars denote photometric redshifts and red squares denote secure public spectroscopic redshifts.  Three secure public spectroscopic redshifts disagree with our multiple line spectroscopic redshifts.  For two of these, we assign a redshift of z$\sim$0.5  versus z$\sim$1 for the public redshifts.  For these two cases, a second spurious line is seen in ROLES data which makes a genuine \oii~emission line appear to be H$\beta$~or \oiii.  For the third discrepant point, the public redshift puts \oii~just outside our redshift window, meaning that we have assigned two spurious lines to \oii~and [Ne\,{\sc iii}] when we should actually have recorded no emission lines.  These multiple spurious lines coinciding with the correct spacings for genuine lines are rare in ROLES as a whole.\footnote{For the few rare instances where a secure public spectroscopic redshift exists which disagrees with a ROLES' redshift, we favour the former value, since the alternate spectroscopy likely has multiple emission lines (due to the longer wavelength coverage than our band-limited data) or other features which make the other redshift more secure than ours.}  For the remaining secure public spectroscopic redshifts, we agree with the public redshift.  Similarly, the vast majority of photometric redshifts agree well with our multiple line spectroscopic redshifts.  A small number of outliers exist, which may be due either to serendipitous combinations of spurious lines, or catastrophic failures of the photometric redshift.  We note that plotting a single best fit photometric redshift and confidence interval is not quite the same as the method we use to interpret the redshift of an emission line (as described below) and, in practice, our weighting scheme will be more robust than this simple comparison.

\subsection{Line identification}
\label{sec:lineid}
For the majority of these emission line objects, only a single
line is detected and thus the spectrum on its own gives an ambiguous redshift.
Thus we must make use of additional data to determine a redshift.  For
a subsample of these objects, spectroscopic redshifts are available from
public spectroscopy.  We again use the compilations from
\citet{Wuyts:2008lq} and \citet{Tran:2007hx}, considering only secure redshifts. 
  For most objects, public spectroscopic redshifts
are not available (only 24 ROLES' galaxies already have spectroscopic redshifts from public spectroscopy, which is not surprising given their faintness) and we must use photometric redshifts in order to determine the most likely identity of our emission line.  Since our
survey began, more accurate public photometric redshifts
for the CDFS have become available \citep[FIREWORKS,][]{Wuyts:2008lq}.  They utilise many more filters and cover a wider wavelength range than those used in the Mobasher \& Dahlen catalogues
and hereafter we adopt these.  For consistency with the original $K$-band selection, we retain the original $K$-band total magnitudes from the Mobasher \& Dahlen catalogue.\footnote{The agreement between total magnitudes is generally very good since they are based on the same original images, but with a slightly different reduction and photometry procedure.  However, a small number of sources from the Mobasher \& Dahlen catalogue are not present in the Wuyts et al. catalogue.  This is presumably due to the former being $R$-band selected and the latter being $K$-band selected, and also the $K$-band surface brightness limit imposed by the latter and/or slightly different masking/deblending parameters used in the object detection.}  For the FIRES field, the photometric redshifts derived in \citet{Forster-Schreiber:2006ih} have been used throughout.  The most probable identities of strong
emission lines in our survey are \oii$\lambda3727$~($0.889 < z < 1.149$), H$\alpha\lambda6563$
($0.073 < z < 0.220$) or one of H$\beta\lambda4861$ or \oiii$\lambda\lambda4959,5007$~($0.406 < z <
0.648$).  Since we only care whether or not the line is \oii~we consider the
likelihood that it is \oii~versus one of the other lines.  In
order to do this, the probability distribution functions
(PDFs) of the photometric redshifts for both fields are used, kindly provided to us by
S. Wuyts.  The photo-z PDF is integrated over the redshift range for
which \oii~could be detected and this is compared with the PDF integrated
over the three different redshift windows just described.  The ratio
then gives the probability that the line is \oii~($P_{\rm OII}$).  If
the probability is $P_{\rm OII}<$0.1 it is set to zero (i.e. assume it
is not \oii) and for $P_{\rm OII}>$0.9 it is set to unity (assume it
is definitely \oii).  The majority of our detections fall into one of
these categories.  

\begin{figure}
{\centering
\includegraphics[width=80mm]{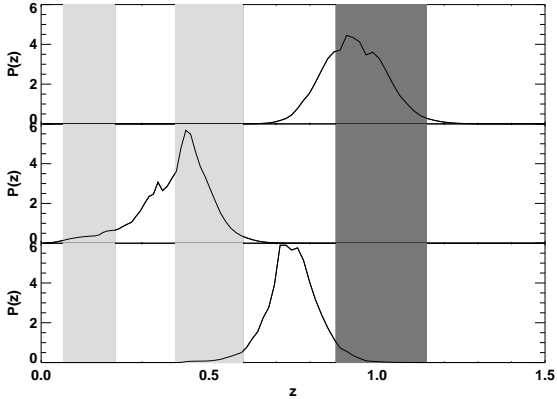}
\caption{Photometric redshift probability distribution functions for three example galaxies.  The probability is in arbitrary units.  Shaded regions show the redshift windows for \oii~(rightmost region, darker shading); $H\beta$/\oiii~(centre shaded region); and H$\alpha$ (lowest redshift region).  The top panel shows a galaxy where the detected emission line would be considered \oii, the centre panel one which would be considered definitely not \oii, and the lower panel shows a galaxy which falls between the two cases and has a probability $P_{\rm OII}\sim0.3$.
\label{fig:pdfs}
}
}
\end{figure}

Of the 246 detected emission line objects in CDFS and 119 in FIRES, only 17 have 0.1$<P_{\rm OII}<$0.9 in CDFS and 40 in FIRES. 195 have $P_{\rm OII}>$0.9 in CDFS and 50 in FIRES. The higher fraction of ambiguous objects in FIRES relative to CDFS is due to the slightly less well-constrained photo-z PDFs of the former due to the different photometric filters available for each field (see \S3.1).  For the few cases where $0.1 < P_{\rm OII} < 0.9$
this weighting is propagated through the analysis such that these
galaxies contribute a fraction $P_{\rm OII}$ of their properties to
the measurement under consideration (i.e. a galaxy with $P_{\rm
OII}=0.5$ would contribute half its \oii~flux to a measurement of the
total \oii~flux).  Fig~\ref{fig:pdfs} shows examples of PDFs for
the three different cases. 

\subsubsection{Flux measurement}
\label{sec:flux}

Line fluxes from the 2D science frames are measured in a 5 $\times$ 5 pixel box around the centre of each emission line.  For a few detections, one row of pixels in this box may fall outside our masked region (i.e., onto bad pixels).  In these cases the profile is corrected by mirroring pixels on the opposite side of the 5$\times$5 box into the bad region.  Flux errors are determined by applying the same method to the noise frame (without convolution).  It is important to distinguish between the detection errors/significances, computed from the optimally-filtered noise frame, and the flux errors, computed from the original (i.e. prior to convolution) noise frame resulting from eqn.~\ref{eqn:noise1}.

To calibrate these fluxes, the ESO spectrophotometric standard star LTT3864 was observed with the same instrumental setup, through one of the slits of the science masks.  To compensate for absorption by
telluric features, we isolate the $O_2$ A-band feature around 7600\AA~ in
our flux standard and measure the ratio of the observed flux to that
expected from the smooth fit.  

\begin{figure}
{\centering
\includegraphics[width=80mm]{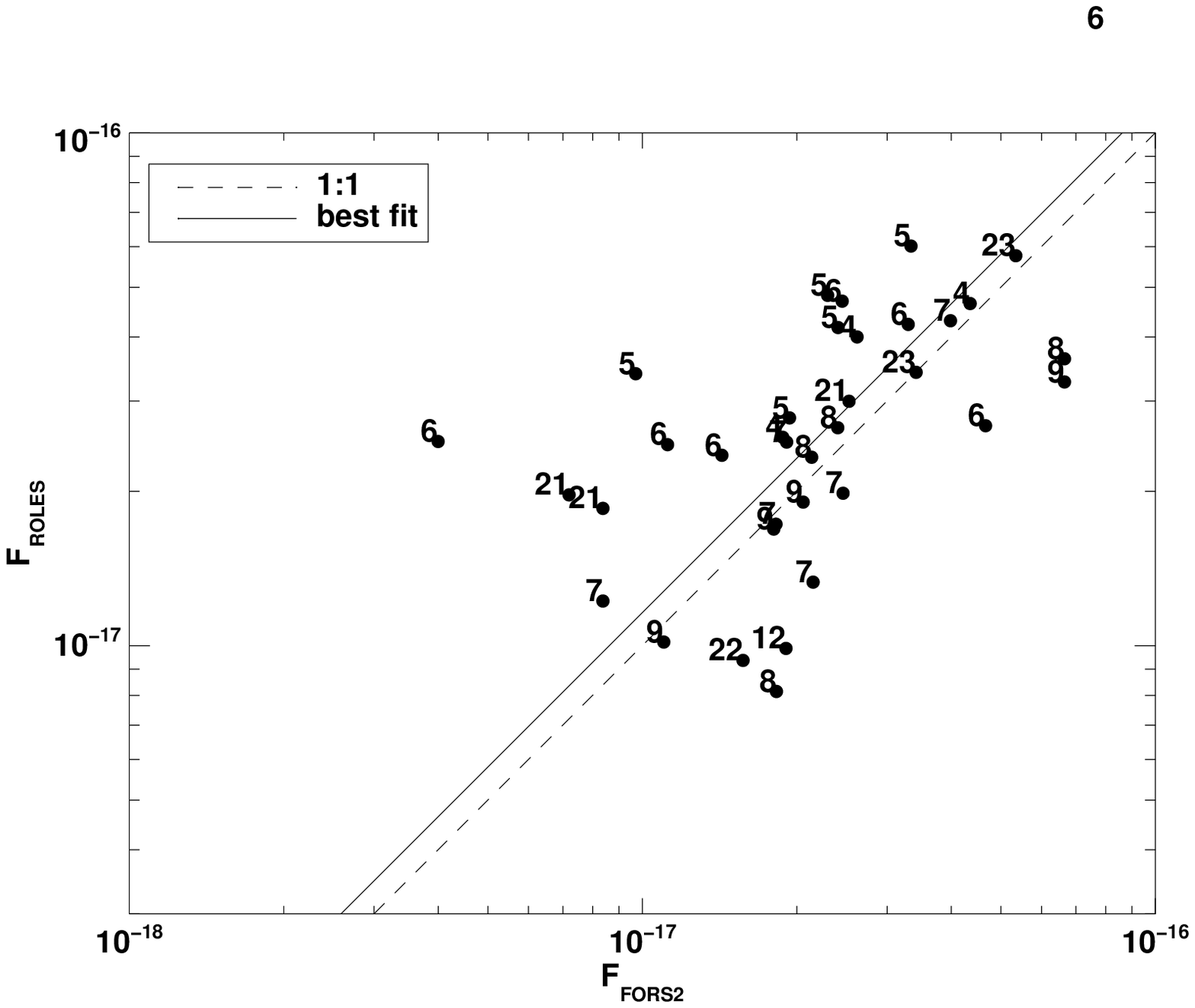}
\caption{Comparison of line fluxes between \citet{Vanzella:2008vq} (FORS2) and ROLES for all objects in common.  Numbers indicate ROLES mask number.  The two biggest outliers lie in the $O_2$ A-band region -- the ROLES fluxes are corrected for telluric absorption, whereas the FORS2 fluxes are not. 
\label{fig:cfflux}
}
}
\end{figure}

\begin{figure}
{\centering
\includegraphics[width=80mm]{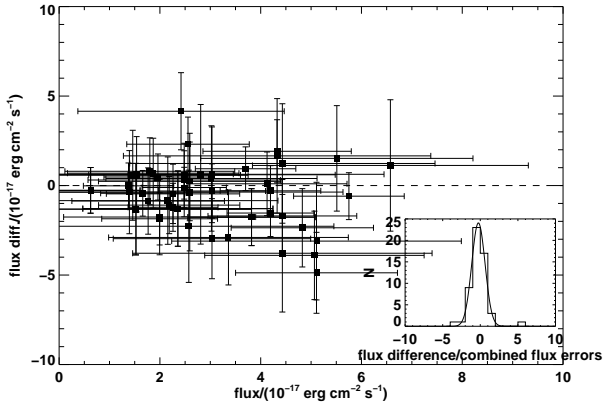}
\caption{Flux difference between repeated measurements of emission lines on different masks.  Inset panel shows the distribution of flux difference divided by the combined flux errors. The overplotted curve is a Gaussian of width 0.9$\sigma$ indicating that our estimates of the flux errors are reasonable.  See text for discussion.     
\label{fig:dflux}
}
}
\end{figure}

The zeropoint of our flux scale is verified using measurements of science targets overlapping with the extensive spectroscopy in the CDFS.  We use ESO public 1D spectra\footnote{see: http://www.eso.org/science/goods/} observed with FORS2 in the CDFS \citep{Vanzella:2008vq}.  Objects in common with our survey are identified and line fluxes are measured using a similar method to that used in ROLES (but on the 1D spectra). The flux calibration of the ESO public data has been extensively checked against multi-band {\it ACS} photometry for a large sample of objects and there does not appear to be anything unusual about the objects in our overlapping subsample (E. Vanzella, priv. comm.).  The \citet{Vanzella:2008vq} spectroscopy is calibrated to reproduce the total flux in a point source.  We have examined the sizes of our ROLES galaxies in {\it ACS} $i$-band images and find that most will have sizes very close to a point source when observed with our typical ground-based seeing ($\sim$0.8\arcsec) and thus \citet{Vanzella:2008vq} should give us approximately total fluxes for our objects.  We plot the flux in our spectroscopy against the public FORS2 spectroscopy in Fig.~\ref{fig:cfflux}.  We find excellent agreement between the two samples, with the scatter dominated by measurement uncertainties.  Error bars have been omitted for clarity, but the size of our errors can be gauged from Fig.~\ref{fig:dflux} and the FORS2 errors are comparable.  The numbers indicate the ROLES mask on which our measurement was made.  There are no obvious systematic offsets with mask number, indicating that this single zeropoint calibration should apply equally well to our whole survey.

The repeatability of our flux measurements is shown in Fig.~\ref{fig:dflux} from repeat observations as described in \S\ref{sec:tests}.  The agreement between independent measurements of the flux in different masks is excellent.  In the inset panel, we compare the errors on each measurement with the combined errors (sum in quadrature) of these repeated observations.  We find that the resulting distribution is approximately Gaussian, with a width of 0.9$\sigma$ indicating that our estimates of the noise are reasonable.

\section{Analysis}
\label{sec:analysis}

\subsection{Stellar masses}
\label{sec:masses}

Stellar masses\footnote{Stellar masses throughout this paper refer to the present mass in stars, not including remnants} are obtained by fitting the extensive multiwavelength photometry at the spectroscopic redshift to a grid of stellar population models (using PEGASE.2, \citealt{Fioc:1997cl}) as described in \citet{Glazebrook:2004zr}. For the CDFS, this consists of $U_{38}, B_{435}, B, V, V_{606}, R, i_{775}, I, z_{850}, J, H, K_s$, [3.6$\mu$m], [4.5$\mu$m], and [8.0$\mu$m] photometry\footnote{In paper I we used a subsample of these filters, not including the {\it Spitzer} data.  As we noted in that paper, omitting the {\it Spitzer} data makes negligible difference to the fitted masses at these redshifts and thus the absence of MIR data for FIRES does not compromise the stellar masses for this field.} (see \citealt{Wuyts:2008lq} for details); and for FIRES this consists of $U, B, V, V_{606}, I_{814}, J_s, H,$ and $K_s$ photometry (see \citealt{Forster-Schreiber:2006ih} for details).  The FIRES catalogue was corrected for Galactic extinction using the maps of \citet{1998ApJ...500..525S}.  For CDFS the extinction, even in the blue, is negligible.  The matched apertures used for measuring colours (see the cited papers for details) are used in the SED-fitting.  Fitted quantities related to total luminosity in the stellar population models, such as stellar mass and SFR, must therefore be scaled from the aperture measurements to total light measurements. This is achieved by multiplying by $10^{-0.4(K_{tot}-K_{ap})}$ where $K_{ap}$ and $K_{tot}$ are the $K$-band aperture magnitude and the total magnitude, respectively.

We note that the colour distributions for the CDFS and FIRES fields are significantly different (both for our spectroscopic sample and for the parent catalogues, although this is consistent with the expectations of cosmic variance for our narrow redshift slice) in the sense that CDFS galaxies are bluer.  This means that for a given $K$-band magnitude, CDFS probes to lower stellar masses than FIRES by around 0.2dex.

\subsection{Nominal L([OII]) Conversion}
\label{sec:loii_conversion}

We use the same conversion of L([OII]) to SFR as J05 which in turn was
based on the calibration of \citet{Kennicutt:1998pa}.  We convert L([OII]) to
L(H$\alpha$) assuming that ([OII]/H$\alpha$)$_{obs}$ = 0.5. We then
correct for extinction in H$\alpha$ assuming average extinction of
A$_{H_{\alpha}}$=1.  Thus the conversion is:

\begin{equation}
	SFR (M_{\odot} yr^{-1}) = \frac{10^{0.4}}{0.5} \times \frac{7.9 \times
	10^{-42}}{1.82} L([OII]) (ergs^{-1}) \label{eqn:sfr_oii}
\end{equation}

\noindent
where the factor of 1.82 accounts for the conversion from a Salpeter
IMF \citep{Salpeter:1955zs} to that of BG03.

\subsection{Spectroscopic completeness}
\label{sec:completeness}

\begin{figure}
{\centering
\includegraphics[width=80mm]{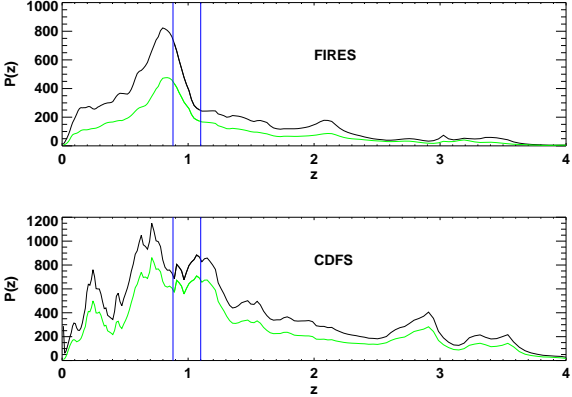}
\caption{Redshift distributions used to calculate the completeness in FIRES (upper panel) CFDS (lower panel).  These are constructed by simply summing the FIREWORKS photometric redshift P(z)'s for individual galaxies (or delta functions for galaxies with spectroscopic redshifts).  Plot shows distributions for all $22.5 < K \le 24.0$ galaxies in our survey area (upper black line) and those on which we placed slits (lower green line).  Vertical lines indicate the redshift limits of our survey.  The completeness is given by the ratio of the integral of the upper line to the lower line within this window.  See text for details. 
\label{fig:pnz}
}
}
\end{figure}

In more traditional spectroscopic surveys, it may be more natural to split the completeness into `targeting completeness' (i.e. the fraction of galaxies of interest on which slits are placed) and `redshift success rate' (the fraction of objects with slits yielding redshift).  Since we use photo-z selection and implicitly assume we obtain redshifts for 100\% of objects within our redshift window with \oii~emission above our flux limit, we simply combine these into a single `completeness' term.  To calculate the completeness of our spectroscopy, we again use the FIREWORKS and FIRES photometric redshift PDFs and public spectroscopy.  For every galaxy in our $K$-selected sample, in the area of our spectroscopic pointing, we sum their PDFs.  This summation gives us the total redshift distribution of all galaxies in the sample (upper curve of Fig.~\ref{fig:pnz}).  The number of galaxies in the $k^{th}$ magnitude bin in the redshift range, $z_0$ to $z_1$ is then

\begin{equation}
\label{eqn:nk}
	N_{k} = \sum_k \int_{z_0}^{z_1} P_k(z) dz ,
\end{equation}

where $P(z)$ is the redshift probability density function given by the photo-z PDF. \footnote{Replacing the photo-z $P(z)$ with a delta function at the spectroscopic redshift, where redshifts are available from public spectroscopy, makes negligible difference to the resulting completeness.}

We then repeat this process for just the galaxies on which we placed slits.  This gives the lower, thicker, green curve in Fig.~\ref{fig:pnz}.  Since we are only interested in a specific redshift range (the shaded region in the figure), the completeness is given by the ratio of the integrals of the PDFs over this redshift range, i.e. the ratio of the area under the upper curve to the lower curve in this redshift range, i.e. 

\begin{equation}
\label{eqn:complete}
	w_k = \frac{N_{slits}}{N_{phot} },
\end{equation}

where $w_k$ is the completeness in the $k^{th}$ magnitude bin, $N_{slits}$ is the number of galaxies on which we placed slits and $N_{phot}$ is the number in the entire photometric catalogue, in the same field, and $N$ are calculated as given in eqn.~\ref{eqn:nk}.

\begin{figure}
{\centering
\includegraphics[width=80mm]{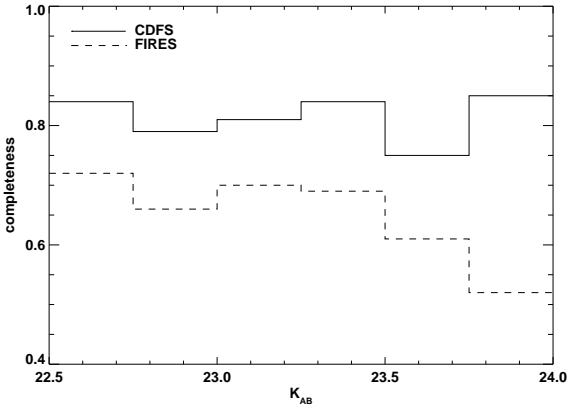}
\caption{Spectroscopic completeness, calculated as described in the text, as a function of $K$-band magnitude for the two ROLES fields. Note that the completeness axis shows down to 40\%, not 0\%.  
\label{fig:completeness}
}
}
\end{figure}

Our overall survey completeness is $\sim$0.85 for CDFS and $\sim$0.7 for FIRES.   In practice, the completeness is a function of $K$-band magnitude, so we repeat this procedure in 0.25 magnitude bins.   For CDFS, our selection has removed this dependence, making it nearly independent of magnitude, but for FIRES there is still a weak magnitude dependence, the completeness dropping to 0.50 in the faintest bin.\footnote{The lower overall completeness for FIRES relative to CDFS for the same observing strategy is due to the presence of the cluster MS1054-03 at z$=$0.83, just outside our redshift window.  Given typical photo-z uncertainties, a significant fraction of objects from this cluster still leak into our redshift selection window (see Fig.~\ref{fig:pnz}).} The results of this are plotted in Fig.~\ref{fig:completeness}.

\subsection{Survey volume}
\label{sec:vol}

In order to calculate the SFRD (\S\ref{sec:sfrd}) we  
require an estimate of the maximum volume, $V_{max}$, from which each galaxy observed in our survey is drawn.  For each galaxy, the value of $V_{max}$ is determined by two limits: an
\oii~line flux limit and a $K$-band flux limit.  $V_{max}$ is the total volume in
which the galaxy could be located and yield an \oii~flux above our
flux limit at the observed wavelength and yield a $K$-band magnitude
between our bright and faint $K$-band limits.  

We calculate $V_{max}$ for the $i^{th}$ galaxy as 

\begin{equation}
	V_{max,i} = \Omega \int_{z_0}^{z_1} F_{vis, i}(z) \frac{dV_c}{dz} dz ,
\end{equation}

where $\Omega$ is the angular area of our survey (107.8  arcmin$^2$  for CDFS and 29.1arcmin$^2$ for FIRES), $dV_c/dz$ is the differential comoving volume (e.g., \citealt{Hogg:1999kl}), and $F_{vis, i}(z)$ is a visibility function for each galaxy which is 1 when both its $K$-band and \oii~fluxes are above our survey limit at that redshift (wavelength), and 0 otherwise.  The calculation of the two terms (based on the $K$-band and \oii~flux limits) making up $F_{vis, i}(z)$ are outlined below.  For reference, a galaxy which is visible at all redshifts (from $z_0=0.889$ to $z_1=1.149$) in ROLES would be drawn from a volume of $7.4 \times 10^4$ Mpc$^3$.

\subsubsection{$K$-band flux limit}

For a galaxy observed
at redshift $z_{obs}$ with $K$-band magnitude $K_{obs}$, 
the $K$-band magnitude at a given redshift $z$ is estimated to be:

\begin{equation}
	K_z = K_{obs} + 5log(d_{obs}/d_{z}) + (k\mbox{-}corr_z - k\mbox{-}corr_{obs})
	\label{eqn:sfrd2}
\end{equation}

\noindent
where $d_{obs}$ and $d_z$ are the luminosity distance at the observed
redshift and the redshift at which we wish to estimate $K$
respectively.  $k$-$corr_{obs}$ and $k$-$corr_{z}$ are the $k$-corrections at
the same redshifts.  For simplicity we adopt the $K$-band differential $k$-correction of \citet{Glazebrook:1995iu}, which is based on an average model SED for normal galaxies: 

\begin{equation}
	k\mbox{-}corr(z) = \frac{-2.58z + 6.67z^2 - 5.73z^3 - 0.42z^4}{1 - 2.36z + 3.82z^2 - 3.53z^3 + 3.35z^4}
	\label{eqn:sfrd3}
\end{equation}

In any case, in the $K$-band the differential $k$-correction depends only slightly on the galaxy type (for z$<$1.5), and due to our narrow redshift slice the $k$-correction is constant to a good approximation.  In practice, the effect of applying a $k$-correction or not only makes a $\sim$1\% difference to the total volume probed.

\subsubsection{{\rm \oii}~flux limit}

\begin{figure}
{\centering
\includegraphics[width=90mm,angle=0]{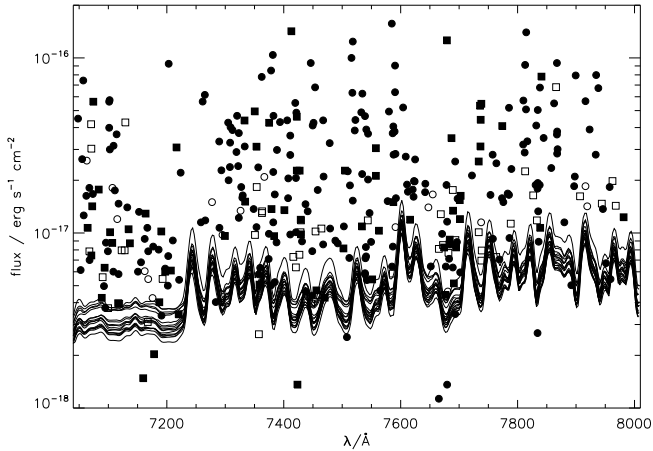}
\caption{Flux versus wavelength for the lines with $P_{OII} >0.5$
(i.e. most-likely \oii~detections, solid circles) and $P_{OII} < 0.5$
(open circles).  The solid lines show the 4$\sigma$ flux limit derived from 4$\times$ the flux measurement error determined from the average noise spectrum for each mask. The limit used in ROLES is that of the shallowest mask (upper line) to be conservative.  Circles show CDFS detections and squares show those from the FIRES field.  Points may appear below the detection threshold, since the flux measurements are made from the unsmoothed data frames; whereas the detection limits refer to the noise level from the Gaussian smoothed frames.  The few points which are above the detection limit but below the measurement limit are rejected from further analysis.
\label{fig:flux_lambda}
}}
\end{figure}

Our \oii~line flux limit is not simply a constant flux limit (as is sometimes assumed in similar surveys), but is in fact a function of wavelength due to the numerous bright night sky lines.  Since we have propagated an estimate of the noise in our spectra, $N_{ij}$, it is a simple matter to calculate the average flux limit as a function of wavelength.  We calculate the average noise spectrum for each mask and then take the shallowest mask as a conservative limit.  The actual value we use is $4 N_{ij}$, since we adopt a 4$\sigma$ detection threshold.  This refers to the flux in the unsmoothed image, which is what is used to make flux measurements.  Since object detection is performed in the gaussian-smoothed images, some detections can appear below the formal flux limit in the unsmoothed frame (as it is only by smoothing to increase the S/N that they pass the detection threshold).  Fig.~\ref{fig:flux_lambda} plots the fluxes of our \oii~lines versus their observed wavelengths. The solid line shows our 4$\sigma$ \oii~limit.  As can be seen, the limit varies from approximately $3\times10^{-18}$ erg s$^{-1}$ cm$^{-2}$ in the deepest regions to around $1 \times 10^{-17}$ erg s$^{-1}$ cm$^{-2}$  underneath the brighter sky lines. Those few points which appear below the flux limit of the shallowest mask (upper solid line) are rejected from further analysis.

Averaged over all of our \oii~lines, calculating the total $V_{max}$ in this way using the detailed noise spectrum gives a value 20\% lower than had we calculated it assuming a constant flux limit of
the flux in our faintest line.

\subsection{Higher mass comparison samples}

In order to extend the mass range considered in this work to higher masses, we employ two external samples.  The first is the Gemini Deep Deep Survey (GDDS).  Details of the GDDS data are given in \citet{2004AJ....127.2455A}.  Briefly, this is a $K$-selected survey at somewhat brighter magnitudes (higher stellar masses) than ROLES, but still much fainter than typically attempted by spectroscopic surveys.  GDDS used N\&S with the GMOS spectrograph on the 8-m Gemini telescope.  Spectroscopic exposure times were typically around 20 hours. Four fields totalling an area of 121 arcmin$^2$ were observed, but the regions selected for spectroscopy were normalised to a larger photometric survey area of 554.7 arcmin$^2$ and so this is the area that should be used when considering sampling volume, cosmic variance, etc.  For the GDDS fields, the photometry available is $VRIzK$, with 2/4 fields lacking $R$-band.  They use colour selection ($I-K$) to preferentially target non-star forming galaxies, but include a sparse ($\sim$1 in 6) sampling of bluer (star-forming) galaxies.  Redshifts are determined interactively, by visually comparing spectra to templates.  This likely yields accurate redshifts where multiple features are visible, but may, for example, not be as highly complete for low SFRs as ROLES (for example, a single 4$\sigma$ emission line in GDDS is less likely to be spotted by eye as in our automated search for this specific class of object).  

For the GDDS data, the photometry was fitted to a similar grid of models as for deriving the ROLES stellar masses and SFRs (\S\ref{sec:masses}).  One minor difference is the dust extinction law used in the model fits (SMC for GDDS vs \citealt{Calzetti:2000sp} for ROLES).  

The second comparison sample comes from the ESO public FORS2 spectroscopy in GOODS-S \citep{Vanzella:2008vq}.  This is largely a $z$-band selected survey, but with some colour-selected (including photometric redshift-selected) subsamples.  This survey uses 1\arcsec~slits and is flux calibrated based on the broadband photometry of continuum sources.  Since this sample covers exactly the same region of sky as ROLES (and thus the same multiwavelength data are available), it provides the ideal sample to extend our survey to higher masses. Our flux calibration is compared with spectroscopy in common with this survey in \S\ref{sec:flux}.  Exposure times with FORS2 (on the 8-m VLT) were typically around 4 hours, and so the spectroscopic data are comparable in limiting \oii~flux, or deeper than the ROLES data.  We take spectroscopic redshifts, confidence flags and 1D flux calibrated spectra\footnote{see http://www.eso.org/science/goods/} for galaxies which directly overlap with our observed LDSS3 pointings in CDFS.  Only the two highest confidence classes of redshifts (A and B) and objects in the ROLES redshift window (0.889 $<$ z $\le$ 1.149) are used.  From the 1D spectra, \oii~luminosities are measured in a way analogous to the ROLES method.  From measurements of many faint emission lines, the 4.0$\sigma$ flux limit is estimated to be $6 \times 10^{-18} erg~s^{-1}~cm^{-2}$.  Objects are matched to our photometry and stellar masses fitted, completeness calculated and \oii~SFR estimated in exactly the same way as for ROLES.  The (targeting) completeness for the FORS2 sample is around 40\% in each $K$-band magnitude bin, except for the brightest bin (highest mass galaxies) which is around 70\% complete.  The success rate of measuring redshifts is taken to be 72\% \citep{Vanzella:2008vq} independent of magnitude (which is likely a reasonable approximation for emission line galaxies).

\section{Results \& discussion}

\subsection{{\rm \oii}~luminosity function}
\begin{figure*}
{\centering
\includegraphics[width=85mm,angle=0]{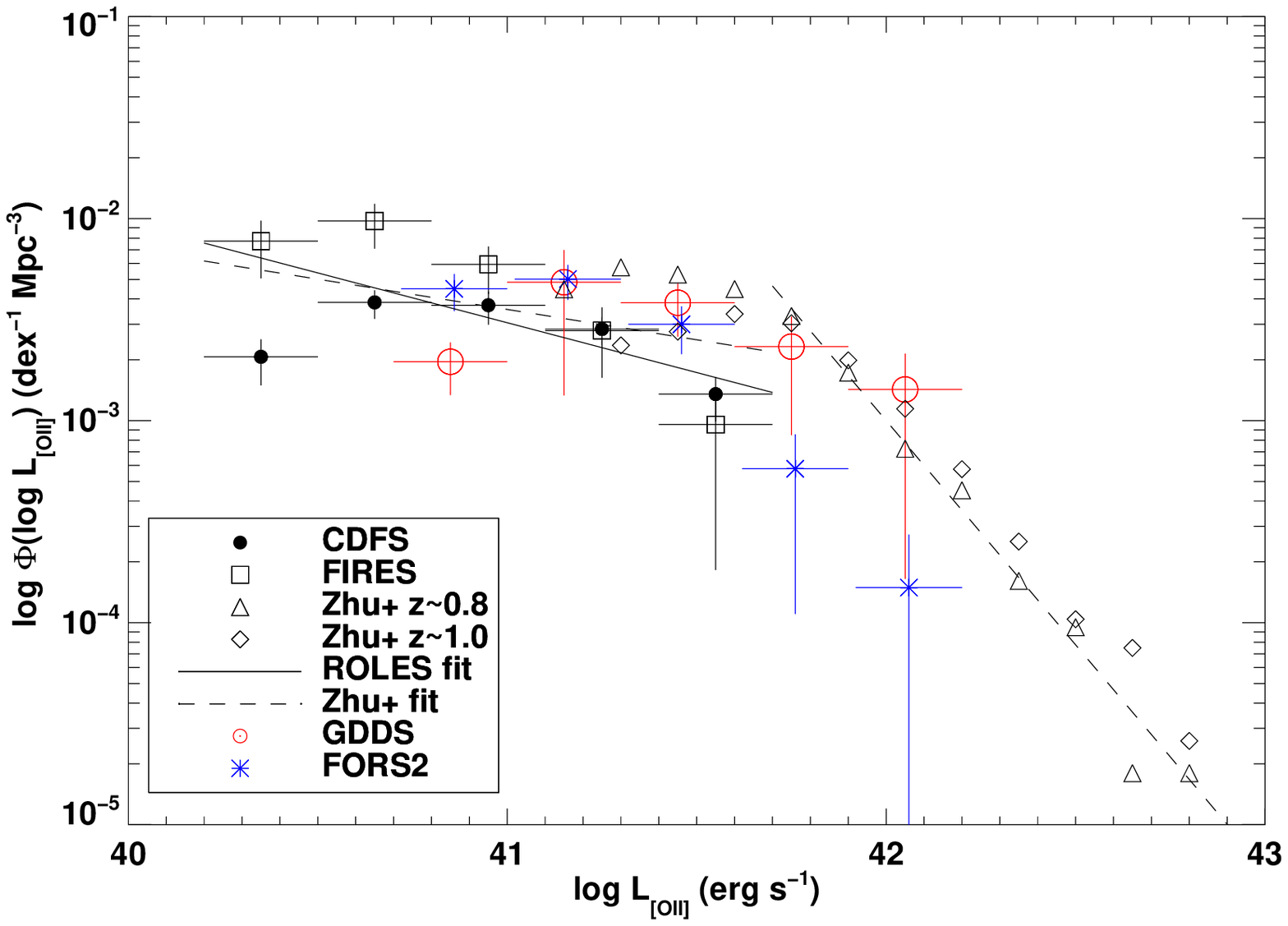}
\includegraphics[width=85mm,angle=0]{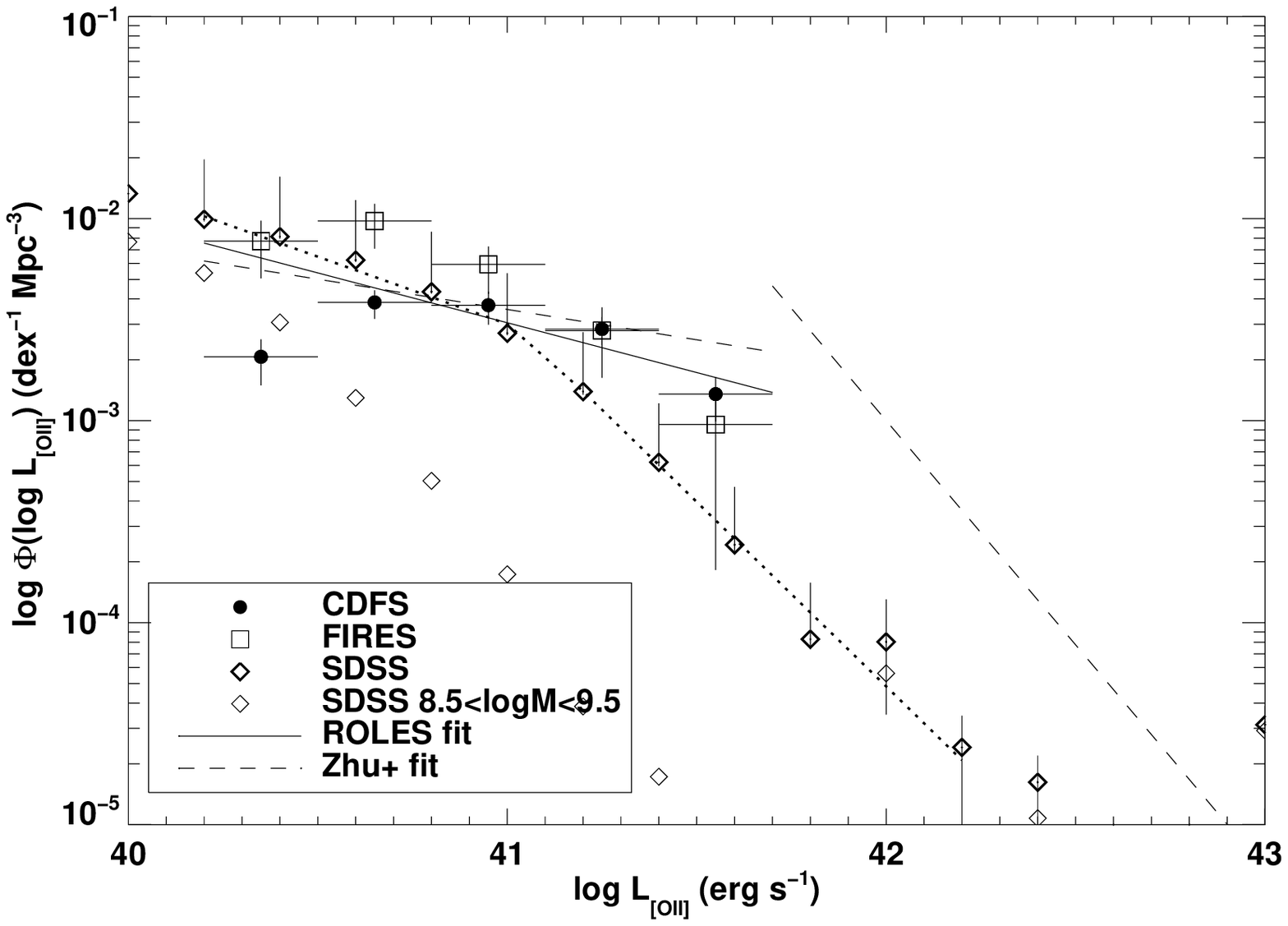}
\caption{\oii~luminosity function.  Filled circles and open squares show measurements made in the CDFS and FIRES fields respectively.  Error bars are derived from Poisson errors only.  In the left panel, open triangles and diamonds show the measurements of \citet{Zhu:2008iv} in their $0.752< z < 0.926$ and $0.926<z<1.099$ bins, respectively.  The dashed lines show the best fit line found by \citet{Zhu:2008iv} for $\log L_{[OII]} \gsim 42$ and their assumed faint end slope of $\alpha=-1.3$ at low luminosities, renormalised to pass through our data.  The thick line shows our best fit faint end slope of $\alpha=-1.5$.  Red open circles and blue asterisks show comparison data from the higher mass samples of GDDS and FORS2 (offset slightly for clarity) respectively. In the right panel, the z$\sim$1 ROLES measurements are now compared with the z$\sim$0.1 measurements made in SDSS (for mass cuts similar to the ROLES' mass range: diamonds; and for no mass cuts: thicker diamonds with error bars).  
\label{fig:oiilf}
}}
\end{figure*}

Fig.~\ref{fig:oiilf} shows the \oii~luminosity function (LF) measured by ROLES.  For the $i^{th}$ bin of log luminosity, the number density, $\Phi_i$,  is
calculated using the 1/{\it Vmax} method as 

\begin{equation}
	\Phi_i = \sum_{i} \frac{P_{OII, i}}{V_{max, i}~w_i}
	\label{eqn:lfvmax}
\end{equation}

\noindent
where $P_{OII}$ is the weighting due to the probability that the
emission line is \oii, $V_{max, i}$ is the maximum volume in which galaxy $i$
could be located and have been found in the ROLES survey, and $w_i$ is the weighting given by the spectroscopic completeness (eqn.~\ref{eqn:complete}).  Error bars are calculated by scaling Poisson distributions following the method of \citet{Zhu:2008iv}, as summarised below.  A scale factor within each log luminosity bin is determined from an effective weight, $W_{eff}$,

\begin{equation}
W_{eff} = \left[\sum_i\frac{1}{(V_{max})_i^2}\right]\Big/\left[\sum_i\frac{1}{(V_{max})_i}\right],
\end{equation}

and the effective number of objects, $N_{eff}$,

\begin{equation}
N_{eff} = \left[\sum_i\frac{1}{(V_{max})_i}\right]\Big/W_{eff}.
\end{equation}
The upper and lower limits are then calculated for the effective number assuming a Poisson distribution (\citealt{Gehrels:1986gd} provides a useful way to approximate this) and these numbers are scaled by $W_{eff}$ to give the upper and lower SFRD limit in each mass bin.

Fig.~\ref{fig:oiilf} shows the measurements for the CDFS and FIRES fields separately (filled circles and open squares respectively). Since cosmic variance is likely to be important, and is not included in the error bars, the difference between the two fields gives an indication of the magnitude of this effect.  Estimating the square root of cosmic variance, $\sigma_V$, from \citet{Somerville:2004ff} using the number density of ROLES galaxies in these fields gives a value of $\sigma_V \sim0.7$.  This is consistent with the observed difference between the two fields.  

Open triangles and diamonds (left panel) show the measurements of \citet{Zhu:2008iv} in their $0.752< z < 0.926$ and $0.926<z<1.099$ bins, respectively (error bars omitted for clarity).  They suggest that the z$\sim$1 \oii~LF is best fitted by a double power law model with a turnover around $\log(L_{[OII]}) \sim42$.  Their sample was insufficiently complete at lower luminosities to constrain the faint end slope, $\alpha$ ($dN/dL \propto L^\alpha$), so they assumed $\alpha=-1.3$.  Their fit and assumed slope are shown as the dashed lines in Fig.~\ref{fig:oiilf}.  The faint end line has been normalised to fit through the ROLES data.  It can be seen that the ROLES data do not overlap with the \citet{Zhu:2008iv} sample brighter than the turnover.  This is not unexpected, since ROLES is designed to be a low stellar mass sample, and thus low mass galaxies cannot sustain the high SFRs implied by such high \oii~luminosities (e.g., \citealt{Noeske:2007tw}). This mass selection is likely to be the biggest difference in comparing the LF between the two samples, and may be responsible for the apparent offset in normalisation where the two samples overlap in luminosity.  We fit a power law to the ROLES data and find a faint end slope of $\alpha \sim-1.5$, although this is not well constrained.  The main purpose of performing this fit is to obtain an indication of the turnover due to incompleteness in \oii~luminosity.  This appears to occur around $\log(L_{[OII]}) \sim40.5$, assuming the intrinsic distribution can be well approximated by a power law.  Using eqn.~\ref{eqn:sfr_oii}, it can be seen that SFR of $1 M_\odot~yr^{-1}$ corresponds to a log luminosity of 40.7, assuming 1 magnitude of extinction at H$\alpha$. This suggests that ROLES is becoming incomplete in SFR at SFR$\approx 0.7 M_\odot~yr^{-1}$.  However, this amount of extinction is likely an overestimate for these low mass galaxies (see e.g., \g09, and next section), and thus the limiting SFR is likely lower.  

In the right panel, the ROLES data are now compared with the z$\sim$0.1 \oii~LF measured in the SDSS (\g09). Open diamonds (error bars omitted for clarity) show the LF for a subsample cut in stellar mass to approximately match the ROLES range ($8.5 < \log(M/M_\odot) < 9.5$), and thicker diamonds show the whole sample of SDSS galaxies, unrestricted in stellar mass.  The mass selection obviously makes a large difference to the shape of the \oii~LF, as mentioned above.  However, if we compare the ROLES data to the SDSS sample unrestricted in mass, we find that the measurements (where the samples overlap, at the faint end) agree within the errors. Assuming at z$\sim$1 ROLES' mass galaxies dominate the \oii~LF in this luminosity range, this is a reasonable comparison.  The fitted faint end slope in the SDSS data is $\alpha_{faint}=-1.67\pm0.02$, compatible with the ROLES fit.  At the bright end in SDSS, $\alpha_{bright}=-2.83\pm0.29$, somewhat shallower (but compatible within the uncertainties) than the z$\sim$1 measurements from \citet{Zhu:2008iv} which are around -3.2 -- -3.0 $\pm$0.1 for the data shown in Fig.~\ref{fig:oiilf}. Although the bright and faint end slopes are consistent with their z$\sim$1 values, the position of the turnover clearly evolves. The measured values are $\log L_{TO}=41.0$ in SDSS (\g09) and $\log L_{TO}=41.9$ at z$\sim$1 \citep{Zhu:2008iv}.   \citet{Zhu:2008iv} propose a parameterisation for the evolution of the bright end of the \oii~LF (which they measure between $0.75\lsim z \lsim1.45$) by considering the the redshift evolution of objects where the number density is $10^{-3.5} dex^{-1} Mpc^{-3}$ as $\log L_{[OII]}=0.46z+41.85$.  Extrapolating this relation to z$\sim$0.1 would give an expected luminosity for this number density of objects of $\log L_{[OII]}=41.9$.  In fact, the value measured from SDSS is closer to $\log L_{[OII]}=41.4$, half a magnitude brighter.  Thus the evolution is stronger between z$\sim$0.1 and z$\sim$1 than that expected from a simple extrapolation of that between $0.7\lsim z \lsim 1.5$. 

Under our simple conversion of \oii~luminosity to SFR, Fig.~\ref{fig:oiilf} can be viewed as a SFR function.  This may suggest that the SFR function has remained constant in shape but has simply shifted toward lower SFR (lower \oii~luminosity) at lower redshifts.  This could imply that galaxies have simply decreased their SFR by the same amount, on average, over this redshift range.  The \oii~LF is a relatively blunt tool for understanding how the SFR of the general population evolves, and so in the next section we examine the star formation rate density as a function of redshift and stellar mass.

\subsection{Star formation rate density}
\label{sec:sfrd}

For each mass bin, the star formation rate density, $\rho_{SFR}$, is
calculated using the 1/$V_{max}$ method as 

\begin{equation}
	\rho_{SFR} = \sum_{i} \frac{P_{OII, i}~SFR_i}{V_{max, i}~w_i}
	\label{eqn:sfrd1}
\end{equation}

\noindent
where the symbols have the same meaning as for eqn.~\ref{eqn:lfvmax} and  $SFR_i$ is the star formation rate (derived from $L_{[{\sc OII}]}$ following eqn.~\ref{eqn:sfr_oii}) of the $i^{th}$ galaxy.  Error bars are again calculated by scaling Poisson errors, as for the \oii~LF, but this time in mass bins instead of luminosity bins.

\begin{figure*}
{\centering
\includegraphics[width=180mm,angle=0]{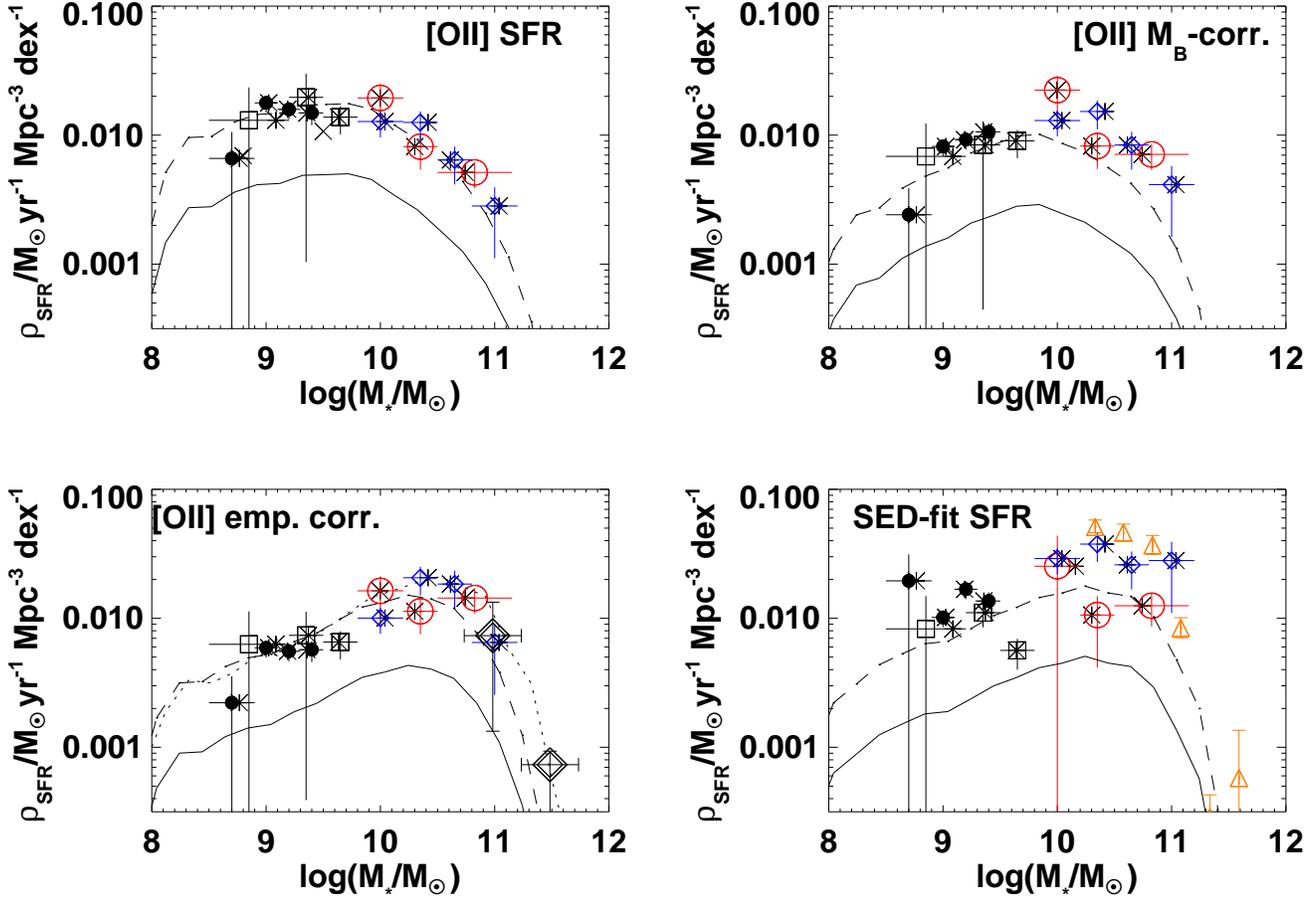}
\caption{The SFRD as a function of stellar mass using various SFR estimators.  In each panel the ROLES data are shown separately for the CDFS and FIRES fields as filled circles and open squares respectively.  (Diagonal crosses show the mean SFRD in the subsample of CDFS from paper 1, for reference.) The higher mass GDDS points are shown as red open circles and FORS2 as blue diamonds.  The equivalent SFR estimator at z$\sim$0.1 taken from SDSS is shown as a solid curve.  The dashed curve is the solid curve renormalised by a factor of 3.5 to agree with the observed z$\sim$1 \oii~SFRD.  The upper left panel shows our default \oii~indicator (constant extinction); upper right panel shows \oii~SFR corrected by observed $B$-band luminosity following the prescription of \citet{Moustakas:2006wp}; lower left panel shows the empirical mass-dependent correction of \g09~to \oii~and the lower right panel shows the SFRD obtained by SED-fitting.  Orange open triangles show the UV-determined SFRD for the spectroscopic sample of \citet{Cowie:2008ob}.  In the lower left panel, the open diamonds show \oii$+$24$\mu$m SFRs from DEEP2 and the dotted line shows the Balmer-decrement corrected H$\alpha$ SFRD (also scaled by a factor of 3.5) from SDSS.  See text for discussion.
\label{fig:sfrd}
}}
\end{figure*}

Again, the results are plotted separately for the two fields to estimate the effect of cosmic variance.   Slightly different-sized mass bins are used in the two fields due to the different sample sizes.  Bin sizes are chosen to give comparable numbers of galaxies in each mass bin.  SFRDs, errors, and the numbers of objects used are tabulated in Table \ref{tab:sfrd}.  We note that the mass cuts imposed remove a few objects from the initial sample summarised in \S2.5 (and used for the LF analysis). For objects passing the flux limit and magnitude cuts, we have 212 in CDFS (17 of these are given fractional weights, i.e., have 0.1$<P_{\rm OII}<$0.9) and 90 in FIRES (40 of these with fractional weights). Once the mass cuts are imposed, these numbers are reduced to 199 for CDFS and 86 for FIRES, split by mass as tabulated in Table~\ref{tab:sfrd}.  For a higher mass comparison sample, we also include data from GDDS and FORS2.   We have recalculated SFRDs for GDDS using their raw SFRs, completenesses and weightings (kindly provided by S. Juneau) in exactly the same way as for ROLES\footnote{For GDDS, we use data in the redshift range 0.88$<$z$\le$1.40. The mass bins are chosen to give comparable numbers of objects in the overlap with ROLES' redshift range (0.88$<$z$\le$1.15) and using data in just this range makes negligible difference to our \oii-based SFRD results.  However, UV SFR estimates are only available at z$>$1.2 (due to their limited very blue photometry) and so we extend the redshift range slightly higher than this to allow comparison of UV SFRDs, in the next section.}.  Results for our default \oii~SFR estimator (Eqn.~\ref{eqn:sfr_oii}) are shown in the top left panel of Fig.~\ref{fig:sfrd}.  This is the same transformation from \oii~luminosity to SFR as used in GDDS (J05). 
Filled circles and open squares show the results for the CDFS and FIRES fields respectively.  The slightly different mass ranges probed by the CDFS and FIRES fields are due to the different intrinsic colour distributions of the fields and the fixed $K$-band limits used, as described in \S\ref{sec:masses}.  Horizontal error bars show the range of the mass bins, with the data points indicating the centre of the adopted bin.  Asterisks indicate the median mass within the bin.  GDDS points are shown as open red circles.  The fact that the GDDS and FORS2 points seem to join smoothly onto the ROLES points at higher masses is a useful independent check of our absolute flux calibration and means that our calibration is likely as good, or better, than our estimated 30\% absolute uncertainty.    For reference, the two diagonal crosses show the results (error bars omitted for clarity) from the smaller subset of ROLES data presented in paper 1, where we first observed the turnover in the SFRD towards these lower mass galaxies.  It can be seen that our much larger dataset now allows a better determination of the shape of the SFRD, and the use of two separate fields allows the effect of cosmic variance to be assessed.

The solid curve shows the same measurement made at z$\sim$0.1 in the SDSS (\g09).  The dashed curve is the z$\sim$0.1 SFRD renormalised by a factor of 3.5 to match the amplitude of the z$\sim$1 data.  It can be seen that this simple renormalisation leads to good agreement between the shape of the z$\sim$1 and z$\sim$0.1 SFRD, suggesting that simple density evolution may be enough to explain the difference in SFRD between the two epochs.

\begin{table}
\caption{Nominal \oii~SFRD in bins of stellar mass.  $M_l$, $M_u$ and $M_m$ refer to the lower and upper limits of the mass bin (1$\sigma$ Poisson error) and the median $\log($mass$)$ within the bin, respectively.  SFRD, $\rho_{SFR}$, is given in units of $10^{-3} M_\odot yr^{-1} Mpc^{-3}$ per dex of stellar mass. Again, subscripts $u$ and $l$ denote upper and lower limits.  $N_{gal}$ gives the number of galaxies in the mass bin (for GDDS, numbers in parentheses refer to the subsample with UV SFRs: $1.20<z\le1.40$ only).}
\label{tab:sfrd}
\begin{tabular}{rrrrrrl}
\hline
$M_l$ & $M_u$ & $M_m$& $\rho_{SFR}$ & $\rho_{SFR,l}$ & $\rho_{SFR,u}$ & $N_{gal}$\\
            &               &               &  $\times10^3$ &$\times10^3$ &$\times10^3$ & \\
\hline
CDFS & & & & & & \\
 8.50 &  8.90 & 8.77 &  6.60 & 14.61 &  2.61 & 49 \\
 8.90 &  9.10 & 9.02 & 17.79 & 20.61 & 15.34 & 55 \\
 9.10 &  9.30 & 9.19 & 15.85 & 18.42 & 13.63 & 53 \\
 9.30 &  9.50 & 9.36 & 15.28 & 18.25 & 12.78 & 42\\
\hline
FIRES & & & & & & \\
 8.50 & 9.20 & 9.09 & 13.04 & 41.59 &  2.63 & 30\\
 9.20 & 9.50 & 9.37 & 19.63 & 38.23 &  9.32 & 33\\
 9.50 & 9.80 & 9.64 & 13.80 & 17.48 & 10.86 & 23\\
\hline
FORS2 & & & & & & \\
  9.80 & 10.20 & 10.04 &  12.75 & 15.87 & 10.21 & 31\\
10.20 & 10.50 & 10.42 & 12.54 & 15.97 &  9.81 & 21\\
10.50 & 10.80 & 10.61 &  6.43  & 8.71  & 4.71 & 15\\
10.80 & 11.20 & 11.04  & 2.83  & 4.54  & 1.71 & 6\\
\hline
\multicolumn{3}{l}{GDDS $(0.88<z\le1.40)$} & & & & \\
 9.80 & 10.20 & 9.96 & 19.43 & 26.68 & 14.02 & 13 (2)\\
10.20 & 10.50 & 10.30 &  8.14 & 10.68 &  6.18 & 17 (6) \\
10.50 & 11.15 & 10.75 &  5.14 &  6.15 &  4.29 & 36 (16)\\
\hline
\end{tabular}
\end{table}

\subsubsection{Limits on the SFRD from ROLES galaxies with no {\rm \oii}~detection}
\label{sec:undet}

So far, only galaxies with emission line detections have been considered in the analysis.  However, galaxies may be present in the ROLES redshift (and mass) window, but since we do not detect an emission line, we can assign neither a redshift nor a SFR.  In order to estimate how much these objects may contribute to the z$\sim$1 SFRD, we perform the following test.  We repeat our analysis for all remaining galaxies with no emission line detections.  We exclude objects known to lie outside our redshift window from secure redshift measurements (from public spectroscopy or our own multiple line detections), but use the photo-z's of all remaining objects to assign a probability, $P_{OII}$, that they lie in the ROLES redshift window. For simplicity, we only do this for the CDFS field. The much greater area and spectroscopic follow-up means that this provides better statistics than the FIRES field, anyway.  This yields 232 additional galaxies with non-zero probabilities of being in the ROLES redshift range.  For  these objects, we take the peak of their photo-z PDF as their true redshift, and measure the 4$\sigma$ upper limit on the flux at the expected position of \oii~at this redshift.  Obviously, due to the size of the typical photo-z error ($\Delta z\approx$0.1, \citealt{Forster-Schreiber:2006ih},\citealt{Wuyts:2008lq}), this location is not an accurate estimate of the position of the expected \oii~line, but provides a representative estimate of the noise and thus an approximate 4$\sigma$ upper limit to the \oii~luminosity.  Repeating this test using random sampling of the photo-z PDF instead of the peak yields similar results.  These upper limits are then added to the CDFS catalogue and propagated through the SFRD analysis.  The final analysis including these limits results in an increased SFRD in each mass bin which approximately coincides with the upper 1$\sigma$ poisson error bar (as plotted in Fig.~\ref{fig:sfrd}) in all cases.  These limits are not shown on the plot to preserve the clarity of the results from the detected galaxies.  Thus, ROLES does not miss significant star formation from galaxies which we do not detect down to our \oii~flux limit.  

Additionally, we should also note that we have not explicitly removed AGN from our  sample.  The exclusion of AGN will only {\it lower} the \oii~SFRD in ROLES.  Examination of the MIR colours of ROLES galaxies and X-ray point sources using the extensive multiwavelength data in CDFS shows that AGN contamination is of extremely minor importance in our survey.

\subsubsection{Empirically correcting the {\rm \oii}~SFR}

So far, a simple constant scaling between \oii~luminosity and SFR has been assumed.  As discussed in \g09, this is too naive and overestimates the SFRD at low masses and underestimates it at high masses.  Assuming the mass dependence of \oii~luminosity on SFR (from metallicity, dust extinction, ionisation parameter, etc.) is the same at z$\sim$1 as z$\sim$0.1, any trends with redshift are still robust, but the overall shape of the SFRD as a function of mass is likely skewed by this simple estimator.

\citet{Moustakas:2006wp} present an empirical method for correcting \oii-SFRs using rest-frame $B$-band luminosity (which is often used, e.g., \citealt{Zhu:2008iv}).  Although \g09~showed that this method underestimates the size of the correction needed at high stellar masses, we consider it for completeness and to allow comparison with other works using this procedure.  Synthesizing rest-frame $B$-band absolute magnitudes, M$_B$, from the SED-fitting procedure used to derive stellar masses (\S\ref{sec:masses}) we apply this correction to our \oii-SFRs for each galaxy and the resulting SFRD is shown in the top right panel of Fig.~\ref{fig:sfrd}.  It can be seen that this correction decreases the SFRD in low stellar mass objects and increases the SFRD in the highest stellar mass objects.  The solid line shows the identical measurement made at z$\sim$0.1 in SDSS (\g09) and the dashed line shows this line renormalised by the same factor (3.5) which gave agreement for the constant luminosity \oii~SFRD.  Using this alternate estimator, all the ROLES points still give approximately the same agreement with the scaled local SFRD as for the default \oii~estimator, however only one of the three GDDS points and one of the four FORS2 points are now consistent with this curve at the 1$\sigma$ level.   

In the lower left panel of Fig.~\ref{fig:sfrd}, we plot the SFRD calculated using \oii~with an empirical mass-dependent correction derived by \g09 to reconcile \oii~SFRs with Balmer decrement-corrected H$\alpha$ SFRs at z$\sim$0.1.  This assumes that the mass-dependence of \oii~luminosity scales in the same way at this redshift as locally.  Again solid and dashed curves show the corresponding SDSS measurement and the same, renormalised by a factor of 3.5, respectively.  A fairer way to make the comparison with the local sample is using the Balmer decrement-corrected \ha~SFRD rather than the \oii~SFRD since, as shown in \g09, the \oii~flux limit in SDSS is not low enough for the \oii-based SFRD to have converged.  Using the SFR functions in mass bins, \g09~showed that ROLES reaches low enough \oii~fluxes for the SFRD to have converged (assuming the shapes of the z$\sim$1 SFR functions in bins of mass follow the same shapes as at z$\sim$0.1).  The dotted line shows the z$\sim$0.1 Balmer decrement-corrected SFRD scaled by 3.5.  This scaled \ha~curve passes through the z$\sim$1 data systematically higher than the \oii~curve, for the reason just mentioned: incompleteness in the z$\sim$0.1 \oii-selected SFRD, which becomes important at higher masses.  For this reason, we choose to renormalise the scale factor between the z$\sim$1 and z$\sim$0.1 (\ha) data.  A factor of 2.6 is a better fit than the 3.5 previously found from the (incomplete z$\sim$0.1) \oii~data.  From now on, we will use this factor when comparing the empirically-corrected \oii~SFRD with the z$\sim$0.1 \ha~SFRD.  This curve will be plotted in the next figure.

The fact that the renormalised local measurement agrees with our z$\sim$1 measurement is largely a restatement of the agreement for our nominal \oii~measurement (upper left panel).  However, since this estimator is our preferred estimator (using \oii) of total SFR\footnote{The SFRD data points may straightforwardly be obtained by applying the \g09~correction ($SFR_{corr} = SFR_0/\{-1.424\,tanh[(\log M/M_\odot)-9.827)/0.572]+1.700\}$) to the ($SFR_0$) values in Table \ref{tab:sfrd}.}, we can now compare with z$\sim$1 ``total" SFRs using \oii$+$24$\mu$m SFRs from \citet[][DEEP2]{Conselice:2007ns}, shown as red open diamonds.  Although the error bars are large and the overlap with the GDDS, FORS2$+$ROLES sample in mass is minimal, the agreement at \ms$\sim$10 is encouraging.  Now comparing with the rescaled z$\sim$0.1 data, we see that the DEEP2 points are both consistent with the scaled local value.

\subsubsection{SED-fit SFRs}

In the lower right panel of Fig.~\ref{fig:sfrd}, we examine the SFRD where the SFR of each galaxy is estimated from SED-fitting of the deep, multiwavelength photometry at each galaxy's spectroscopic redshift (as used for the stellar mass-fitting in \S\ref{sec:masses}).  This method is now completely independent of \oii~luminosity (except that for ROLES, a galaxy must have a minimum \oii~luminosity to be selected in the survey).  This is largely equivalent to a rest-frame UV luminosity SFR (with the longer wavelength information constraining the contribution of dust extinction).  Hence, for the local SDSS comparison sample, we use the $u$-band SFRD with the dust correction estimated from the Balmer decrement (\g09).  Again, the solid line shows this, and the dashed line is the same renormalised by 3.5 in number density.  At low stellar masses (ROLES), the ``UV" and \oii~SFRDs approximately agree (with the scaled local prediction and hence with each other), albeit with larger scatter compared with \oii, and some suggestion of a systematic offset, since all but one of the SED-fit points lie above the line.  The larger scatter is likely due to the significant photometric errors on such faint objects.  It should be noted that in order to obtain UV SFR estimates for GDDS, the redshift range is restricted to z$>$1.2.  In order to minimise the effect of evolution (which becomes significant between z$\sim$1.0 and z$\sim$1.6, see fig.~2 of J05), we restrict the redshift range to $1.2<z\le1.4$.  This results in very small samples of galaxies in the two lowest GDDS mass bins (2 and 6) and a modest sample of 16 in the highest mass bin.  No such extension of the redshift range is necessary for the FORS2 data, and this sample covers exactly the same redshift range as ROLES, with many more object in FORS2 possessing UV photometry than in GDDS (20-30 vs 2-6).

For comparison, open orange triangles show estimates of the extinction-corrected UV SFR from the spectroscopic redshift (0.9$<z\le$1.5) survey of \citet{Cowie:2008ob}, transformed to our IMF.  These UV SFRD estimates agree well with the SED-fit SFRD from the FORS2 data. For the \ms$\sim$11 mass bin where the GDDS sample size is reasonable, the agreement is also good.  The lower mass GDDS bins measure significantly lower SFRDs than  \citet{Cowie:2008ob}.  Some caution is required when comparing to the  \citet{Cowie:2008ob} data, since their UV SFR is not measured in exactly the same way as in ROLES or GDDS.   \citet{Cowie:2008ob} derive a slightly different calibration of the UV SFR than that used here.  They also use NIR luminosity to estimate the contribution of  older stellar populations to the UV luminosity and subtract this before converting to SFR.  Given that  \citet{Cowie:2008ob} derive these corrections to the UV SFR self-consistently within their data, we do not attempt to correct for the slightly different overall normalisation they use in UV luminosity--SFR, and we just apply a correction for the change from their Salpeter to our BG03 IMF.  Undoing both of these effects would only increase the \citet{Cowie:2008ob} UV SFRD, strengthening the excess seen at higher masses relative to the GDDS data.  
 In light of the fact that the sample size in these bins is tiny for GDDS, and that the FORS2 data (based on much larger numbers of objects) give better agreement with the \citealt{Cowie:2008ob} results, we feel justified in rejecting the GDDS SED-fit SFRD values from further analysis.

\subsubsection{Best estimates of the z$\sim$1 SFRD}

\begin{figure}
{\centering
\includegraphics[width=90mm,angle=0]{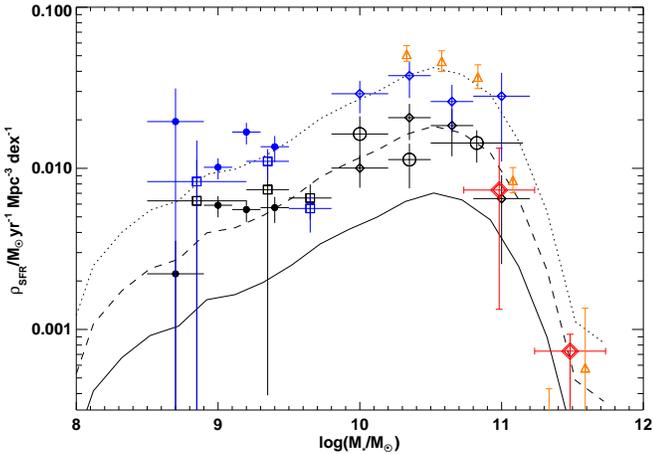}
\caption{Best estimates of SFRD. Symbols indicate the different surveys, as for Fig.~\ref{fig:sfrd}.  Colours refer to the method of estimating SFR: black symbols indicate \oii-based measurements, blue symbols SED-fit SFRs, orange symbols \citet{Cowie:2008ob}'s UV measurements, and red points are for the DEEP2 \oii$+24\mu$m data.  The solid line shows the z$\sim$0.1 Balmer-decrement corrected \ha~SFRD, the dashed line is this scaled by 2.6, and the dotted line this scaled by 6.  The UV/SED-fit based SFR indictors predict a slightly higher overall normalisation for the z$\sim$1 SFRD than the \oii~estimates, but the overall shapes are the same.  See text for discussion.
\label{fig:compsfrd}
}}
\end{figure}

After discussing the various different SFR indicators which may be used with these datasets, it is useful to collate all the results of the best estimators. Our preferred estimators are the empirically-corrected \oii~estimator of \g09, and the SED-fit SFR estimates (after dropping the GDDS results).  These are compiled into one plot in Fig.~\ref{fig:compsfrd}.  Symbols indicate the different surveys, as for Fig.~\ref{fig:sfrd}.  The colours refer to the method of estimating SFR: black symbols indicate \oii-based measurements, blue symbols SED-fit SFRs, orange symbols \citet{Cowie:2008ob}'s UV measurements, and red points are for the DEEP2 \oii$+24\mu$m data.  The solid line shows the z$\sim$0.1 Balmer-decrement corrected \ha~SFRD and the dashed line is this scaled by 2.6. Viewing all this data together, colour-coded by SFR estimator, it is now clear that although the \oii-based SFRDs seem consistent with a renormalisation of $\approx$2.6 times higher than the local SFRD, the SED-fit and UV-based estimates are systematically somewhat higher.  The dotted line shows the local SFRD renormalised by a factor of 6, which is closer to the z$\sim$1 value for the UV/SED results.

Both datasets (\oii~and UV) show comparable shapes for the mass-dependent SFRD - shapes similar to that determined at z$\sim$0.1 from \ha~measurements - but the different estimators disagree somewhat in the overall normalisation for this curve.  We briefly discuss reasons for this possible disagreement (although it should be noted that uncertainties between SFRDs measured with different indicators at these redshifts are typically as large or larger than we see here, e.g., \citealt{Hopkins:2006bv}).  

{\it UV estimate methodology -} \g09~compared UV ($u$-band and FUV)-based estimates of the local SFRD with those of \oii~and \ha~and found that they gave comparable estimates (certainly not discrepant systematically in normalisation by the factor of  $\approx1.7$ seen here).  Although the method of SED-fititng used here differs somewhat from the dust-corrected rest-frame UV luminosities used in \g09, the \citet{Cowie:2008ob} data use a very similar approach to estimating UV SFRs as that used in \g09.  So, the fact that the \citet{Cowie:2008ob} and SED-fit methods give comparable SFRDs suggests that this fitting approach is not responsible for the differences between \oii~and UV SFRDs at z$\sim$1. 

{\it Aperture corrections -} Both the photometric (UV/SED-fit SFR) and emission line (\oii-SFR) estimates require aperture corrections, as discussed in \S\ref{sec:flux}, \S\ref{sec:masses}, respectively.  The former use the $K$-band light scaled from the smaller colour aperture to the value used for the {\it total} magnitude; and the latter use the calibration of \citet{Vanzella:2008vq}, assuming our z$\sim$1 galaxies are seeing-dominated in our spectroscopy.  Again, we can use the fact that the data measured using our method (e.g., the FORS2 data using our photometry and SED-fitting technique compared with the \citealt{Cowie:2008ob} data) agree well with other results in the same mass range as a check that our aperture corrections cannot be egregious.  Similarly, the agreement between the GDDS \oii-SFRD and our \oii-SFRD (for the highest mass galaxies where the aperture corrections should be the largest) shows that uncertainties in the aperture corrections are relatively unimportant, since GDDS do not explicitly use an aperture correction and for the FORS2 data we used the correction based on ROLES data.

{\it Incompleteness in {\rm \oii} SFRD - } It is worth emphasizing that the lower overall SFRD measured from \oii~versus UV is unlikely to be due to missing flux below the \oii~limits of the various surveys.  As we have estimated in \S\ref{sec:undet}, galaxies in ROLES with no detectable \oii~emission, but plausibly within our redshift window would, at most, move the \oii~SFRD to the top of our 1$\sigma$ SFRD errors, causing better agreement between \oii~and UV SFRDs.  However, two important points must be borne in mind.  Firstly, this only applies for the ROLES dataset, which is \oii-selected.  For FORS2 and GDDS, no such explicit dependence on \oii-selection applies.  Secondly, since ROLES is \oii-selected, including more objects in the \oii-SFRD would also increase the contribution to the UV SFRD, since these new entries would then have their SEDs fitted in the same way as for the galaxies currently in ROLES, driving up the UV SFRD.  Additionally, the checks done in \g09~show that the SFRD has likely already converged above the ROLES' (and other surveys employed here) flux limit.  

{\it IMF -} The \oii~and UV luminosity trace SFR via their dependence on the presence of high mass stars.  However, they are sensitive to stars in different mass ranges: $\gsim10 M_\odot$ for \oii~and $\gsim5 M_\odot$ for UV (e.g., \citealt{Kennicutt:1998pa}).  Thus, a different slope of the stellar IMF can lead to a different normalisation between the two estimators.  The agreement between the estimators at z$\sim$0.1 would mean that, if the IMF is responsible, then it would have to be an evolutionary effect (which affects galaxies of all masses approximately equally).  An evolving IMF has been suggested as a way to reconcile estimates of the growth of stellar mass with the cosmic star formation rate history (e.g., \citealt{Dave:2008kl} - an evolving high mass turnover which moves towards lower masses at higher redshift).  Another possibility is that instead of a universal IMF which evolves, the IMF is different (but constant with time) for starbursts versus more quiescent star formation.  There is good motivation for assuming that the IMF in starbursts may be relatively top heavy (e.g., \citealt{Lacey:2008pi} and references therein).  Then, the evolving contribution of starbursts to the total SFRD would lead to an evolution of the relative normalisation of the indicators.  It should be noted that, if the incidence of starbursts is higher at higher redshift, then an increasing contribution from a top-heavy IMF at higher redshift would go in the wrong direction.  i.e., one would expect that the indicator more sensitive to the most massive stars (\oii) would show a higher SFRD than the other (UV), under the assumption of the incorrect IMF.  Conversely, the \citet{Dave:2008kl} model for IMF evolution goes in the correct direction to explain our data.

{\it Evolution of the empirical {\rm \oii}~correction -} We have corrected our SFR based on \oii~luminosity as a function of mass using the local empirical calibration of \g09. This accounts for the (mass-dependent) trends of \oii~luminosity--SFR on metallicity, dust and ionisation parameter.  If the z$\sim$1 trends with mass are still comparable for all of these parameters, then the relation is still valid.  However, it is expected that parameters such as dust and average metallicity will evolve with cosmic time.  For example, for the higher mass galaxies in GDDS, \citet{Savaglio:2005hp} found that the metallicity of \ms$\sim$9.5 galaxies at z$\sim$0.7 is $\sim$0.1 dex lower than local galaxies of the same mass.  If metallicity evolution were responsible, differential evolution as a function of mass \citep{Savaglio:2005hp} should lead to a differential correction as a function of mass, which would change the shape of the SFRD.  Of course it is possible that some conspiracy between dust and metallicity evolution changes the normalisation but not the shape. The metallicity dependence of \oii~\citep{Kewley:2004vs} is such that the predicted correction to the \oii-SFR would be {\it lower} for a decreased metallicity, i.e., that our \oii~SFRD should be overestimated relative to the UV, if this is the cause.   In practice, we have corrected empirically for not just the metallicity, but also the dust extinction, etc.  Assessing the validity of this empirical correction at z$\sim$1 will have to await further follow-up data, such as \ha/H$\beta$~spectroscopy of ROLES galaxies, and such work is underway.  Even if the empirical correction is not correct in detail at z$\sim$1, applying this correction is still better than not applying any mass-dependent correction.  The fact that the shape of the \oii-SFRD is similar to that of the UV SFRD, suggests that evolution in the overall correction might not be responsible.

We have attempted to carefully unify measurements of the SFRD from different surveys (which probe different mass ranges) at the same epoch (z$\sim$1), in an attempt to study trends with stellar mass.  Similar caution must be taken when comparing different surveys at different redshifts in order to study evolution.  It is worth noting that we now disagree with the result of J05 who found that their intermediate mass bin ($10.2< $ \ms $ <10.8$) reached the local value of the SFRD by z$\sim$1.  Our result using their data shows that the SFRD for this mass of galaxies is at least 2.6 times the local value.  This difference is because J05 used the \ha~data from \citet{Brinchmann:2004ct} to estimate the z$\sim$0 SFRD and thus has a very different shape from the \oii~SFRD\footnote{ Also compare the shapes of the dotted curve in the lower left panel of Fig.~\ref{fig:sfrd} with the dashed curve in the upper left panel.  These may be divided by 2.6 to return to the original local value. J05 actually used a constant-extinction correction to \ha~to be more consistent with their higher redshift correction to \oii, but the fact that the \oii~and \ha~SFRDs are such intrinsically different shapes is by far the dominant effect.} (\g09),  scaled to their IMF.  As can be seen from comparing the top left panel of Fig.~\ref{fig:sfrd} (which is the \oii~estimator J05 use) with the lower left panel (see also \g09), the local SFRD (solid curves) derived using the different estimators is significantly different in the GDDS mass range.

\section{Conclusions}

We describe the methodology for ROLES, a survey for dwarf galaxies at z$\sim$1.  Using \oii$\lambda$3727 we have estimated the star formation rates (SFRs) of galaxies with spectroscopic redshifts, and masses in the range $8.5 \lsim $\ms$ \lsim 9.5$ down to a limiting SFR of $\sim0.3 M_\odot\,yr^{-1}$.  

We examine the \oii~luminosity function and find a faint end slope of $\alpha_{faint} \sim -1.5$, comparable to the local measurement from SDSS.  This matches on well to the more \oii-luminous (higher mass) sample from the Gemini Deep Deep Survey (GDDS) and ESO public spectroscopy in the GOODS-S field.

By carefully combining the ROLES data with other spectroscopic surveys, we study the mass dependence of the star formation rate density (SFRD) at z$\sim$1 using two independent SFR indicators: empirically-corrected \oii~(using the \g09~prescription), and SED-fitting, which is largely based on rest-frame UV luminosity (for galaxies with spectroscopic redshifts).  By comparing with the local SFRD from Stripe 82 of the SDSS (\g09), we find that both indicators show that the SFRD decreases equally for galaxies of all masses (8.5$\lsim$\ms$\lsim$9.5) between z$\sim$1 and z$\sim$0.1.  The exact change in normalisation depends on the indicator used, with the \oii-based estimate showing a change of a factor of $\approx$2.6 and the UV-based a factor of $\approx$6.  We discuss possible reasons for the discrepancy in normalisation between the indicators, but note that the magnitude of this uncertainty is comparable to the discrepancy between indicators seen in other z$\sim$1 works.  Our result that the shape of the SFRD as a function of stellar mass (and hence the mass range of galaxies dominating the SFRD) does not evolve between z$\sim$1 and z$\sim$0.1 is robust to the choice of indicator.  

The term {\it cosmic downsizing} has been employed to describe various different aspects of the cosmic star-formation history as a function of galactic mass.  In perhaps the most common usage, one might expect that the downsizing picture would describe a situation in which the peak of the SFRD as a function of mass shifts towards lower masses at lower redshifts.  This is contrary to the scenario seen here, in which the shape of the SFRD does not evolve and the normalisation changes equally for all galaxy masses.  However, it is worth keeping in mind that, considering galaxies with high SFRs, these are primarily high mass galaxies (e.g., \citealt{Noeske:2007tw}) and these will indeed both have dominated the SFRD and been more numerous at z$\sim$1.  This could be considered `downsizing'.  In order to avoid the ambiguities in the term `downsizing', we prefer to simply talk in terms of the shape of the SFRD--mass here. In the next paper in the series, we will examine the SFRs and stellar masses of individual galaxies and confront our observations with the latest theoretical models of galaxy formation.

\section*{Acknowledgments}

This paper includes data gathered with the 6.5-m Magellan Telescopes located at Las Campanas Observatory, Chile.  We thank LCO and the OCIW for the allocation of time to this project as part of the LDSS3 instrument project.  We thank T. Dahlen,  H. Flores, S. Juneau, B. Mobasher, C. Ravikumar, E. Vanzella and S. Wuyts  for graciously providing data and for useful discussions, and Hsiao-Wen Chen and Sandra Savaglio for helpful comments.  Karl Glazebrook and I-H Li acknowledge financial support from Australian Research Council (ARC) Discovery Project DP0774469. Karl Glazebrook and Ivan Baldry acknowledge support from the David and Lucille Packard Foundation. MLB acknowledges support from the province of Ontario in the form of an Early Researcher Award.

\label{lastpage}

\end{document}